\documentclass[twocolumn,journal]{IEEEtran}
\usepackage[T1]{fontenc}
\usepackage[latin9]{inputenc}
\usepackage{amsmath}
\usepackage{amssymb}
\usepackage{graphicx}
\usepackage[unicode=true,
 bookmarks=true,bookmarksnumbered=true,bookmarksopen=true,bookmarksopenlevel=1,
 breaklinks=false,pdfborder={0 0 0},pdfborderstyle={},backref=false,colorlinks=false]
 {hyperref}
\hypersetup{pdftitle={Your Title},
 pdfauthor={Your Name},
 pdfpagelayout=OneColumn, pdfnewwindow=true, pdfstartview=XYZ, plainpages=false}

\makeatletter

\newcommand{\lyxdot}{.}

\let\oldforeign@language\foreign@language
\DeclareRobustCommand{\foreign@language}[1]{%
  \lowercase{\oldforeign@language{#1}}}

\usepackage[caption=false,font=footnotesize]{subfig}
\usepackage{cite}
\newcommand{\GG}[1]{} 

\@ifundefined{showcaptionsetup}{}{%
 \PassOptionsToPackage{caption=false}{subfig}}
\usepackage{subfig}
\makeatother

\begin{document}
\title{Distributionally Consistent Simulation of Naturalistic Driving Environment
for Autonomous Vehicle Testing}
\author{Xintao~Yan, Shuo~Feng,~\IEEEmembership{Member,~IEEE}, Haowei~Sun, Henry~X.~Liu,~\IEEEmembership{Member,~IEEE}\thanks{This research was partially funded by
the US Department of Transportation (USDOT) Region 5 University Transportation
Center: Center for Connected and Automated Transportation (CCAT) of
the University of Michigan. \emph{(Corresponding author: Shuo Feng.)}}\thanks{Xintao~Yan and Haowei Sun are with the Department of Civil and Environmental
Engineering, University of Michigan, Ann Arbor, MI 48109 USA, (e-mail:
\protect\href{http://xintaoy@umich.edu}{xintaoy@umich.edu}; \protect\href{http://haoweis@umich.edu}{haoweis@umich.edu}).}\thanks{Shuo~Feng is with the University of Michigan Transportation Research
Institute, Ann Arbor, MI 48109 USA, (e-mail: \protect\href{http://fshuo@umich.edu}{fshuo@umich.edu}).}\thanks{Henry~X.~Liu is with the Department of Civil and Environmental Engineering, MCity, and also University
of Michigan Transportation Research Institute, University of Michigan, Ann Arbor, MI 48109 USA, (e-mail: \protect\href{http://henryliu@umich.edu}{henryliu@umich.edu}).}}
\markboth{}{}
\maketitle
\begin{abstract}
Microscopic traffic simulation provides a controllable, repeatable,
and efficient testing environment for autonomous vehicles (AVs). To
evaluate AVs' safety performance unbiasedly, the probability distributions
of environment statistics in the simulated naturalistic driving environment
(NDE) need to be consistent with those from the real-world driving
environment. However, although human driving behaviors have been extensively
investigated in the transportation engineering field, most existing
models were developed for traffic flow analysis without considering
the distributional consistency of driving behaviors, which could cause
significant evaluation biasedness for AV testing. To fill this research
gap, a distributionally consistent NDE modeling framework is proposed in this paper. Using large-scale naturalistic driving data, empirical distributions
are obtained to construct the stochastic human driving behavior models
under different conditions.
To address the error accumulation problem during the simulation, an optimization-based
method is further designed to refine the empirical behavior models. Specifically,
the vehicle state evolution is modeled as a Markov chain and its stationary
distribution is twisted to match the distribution from the real-world
driving environment. The framework is evaluated in the case study of a multi-lane highway driving simulation, where the distributional accuracy of the generated NDE is validated and the safety performance of an AV model is effectively evaluated.
\end{abstract}

\begin{IEEEkeywords}
Autonomous vehicle, driving behavior, naturalistic driving environment,
simulation, testing and evaluation
\end{IEEEkeywords}

\section{Introduction}

\IEEEPARstart{T}{esting} and evaluation is a critical step in the
development and deployment of autonomous vehicles (AVs), which has
received extensive attention from both the industry and academia in
recent years \cite{kalra2016-Rand-report,zhao2016accelerated,hungar2017test-Pegasus-testing-cite,koren2018adaptive-testing-cite,o2018scalable-testing-cite,li2019parallel-testing-cite,feng2020part1,feng2020part2,feng2020-part3,feng2020-AAP,klischat2020synthesizing-testing-cite,waymo2020-Waymo-safety-report,feng2021intelligent-NADE,sun2021corner-corner-case}.
Prevailing methods test AVs in the naturalistic driving environment
(NDE), observe their performance, and make statistical comparisons
to human driving performance \cite{feng2021intelligent-NADE}. Due
to the rareness of safety-critical events, however, it has been pointed
out that hundreds of millions and sometimes hundreds of billions of
miles would be required in NDE to demonstrate the safety performance of AVs
at the level of human-driven vehicles, which is intolerably inefficient for the real-world testing \cite{kalra2016-Rand-report}.
Therefore, testing AVs in NDE simulations has
attracted increasing attention because of the advantages of controllability,
repeatability, and efficiency \cite{li2019aads-Baidu-AADS,feng2021intelligent-NADE}.

The key to simulation testing is the trustworthiness of the testing
results. As pointed out in various domains \cite{peng2018sim-sim-to-real-gap1,li2019aads-Baidu-AADS,james2019sim-sim-to-real-gap2,zhao2020sim-sim-to-real-gap3},
the simulation-to-reality gap could hinder and even mislead the training
and testing process of an agent. To fill this gap, existing studies
have paid much attention to the fidelity of vehicle dynamics, sensor
models, and photorealistic images based on techniques such as computer
graphics, physics-based modeling, and data
augmentation \cite{li2019aads-Baidu-AADS,manivasagam2020lidarsim,chen2021geosim}.
However, how to model the naturalistic behavior of human-driven vehicles with high fidelity still remains an open question.
To answer this question, many AV
companies tried to replay human driving behaviors according to the
logged data collected from the real-world driving environment. However,
as the human driving behaviors are pre-determined in the logged data,
they cannot interact with AV models, which severely
limits the scenarios that can be simulated. To address this issue,
the human driving models developed in the transportation engineering
field have been applied, such as the Intelligent Driving Model (IDM)
\cite{treiber2000-IDM} and MOBIL \cite{kesting2007-MOBIL} models.
However, although these models can interact with AV models, they were
designed for traffic flow analysis purposes such as reproducing traffic oscillations and the fundamental diagram, which are not suitable for the AV simulation testing (see Section \ref{sec:Related-Work} for more details of related studies).

\begin{figure*}[tp]
\includegraphics[width=1\textwidth]{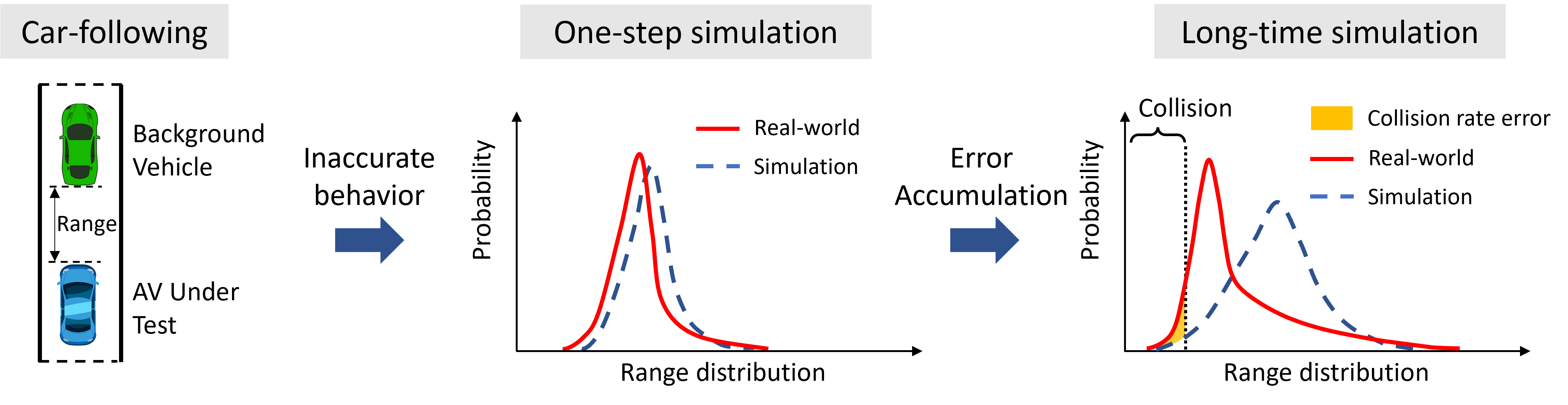}
\caption{Illustration of the distribution inconsistency that might mislead AV development and deployment. \label{fig:mislead-AV-development}}
\end{figure*}

The AV simulation testing brings brand new requirements for the NDE
modeling. To evaluate AVs\textquoteright{} safety performance quantitatively,
the accident rate of AVs in NDE is usually utilized \cite{zhao2016accelerated,feng2020part1,feng2020part2,feng2021intelligent-NADE}.
As the human driving behaviors significantly affect the response and
performance of AVs, estimating the accident rate accurately requires
the NDE model in simulation to be distributionally consistent with
the real-world driving environment. Specifically, let $X$ denote
the variables that define the environment, such as the behaviors of
human drivers. Then, testing an AV in NDE is essential to sample
$X$ from its underlying distribution, denoted as $X\sim P(X)$, to
estimate its performance $\mu_{E}^{\kappa}$ by
\begin{equation}
\mu_{E}^{\kappa}\mathop{:=}\mathbb{E}_{X}\left(\phi_{E}^{\kappa}\left(X\right)\right)\approx\frac{1}{n}\sum_{i=1}^{n}\phi_{E}^{\kappa}\left(X_{i}\right)\approx\frac{m}{n},X_{i}\sim P\left(X\right),\label{eq:AV-performance-expected-probability}
\end{equation}
where $E$ denotes the event of interest (e.g., crash), $\kappa$ denotes the AV
agent under test, $\phi_{E}^{\kappa}\left(X\right)$ denotes the AV
performance at the environment specified by $X$, $n$ denotes the
number of tests, and $m$ denotes the number of event $E$ occurred
during tests. According to the Monte Carlo method \cite{mcbook},
only with the accurate distribution $P(X)$ in the simulation, the
estimation of $\mu_{E}^{\kappa}$ can be statistically accurate. Therefore,
the AV simulation testing requires distributionally consistent NDE
models, which are significantly different from those for traffic flow
analysis. 

\begin{figure*}[tp]
\includegraphics[width=1\textwidth]{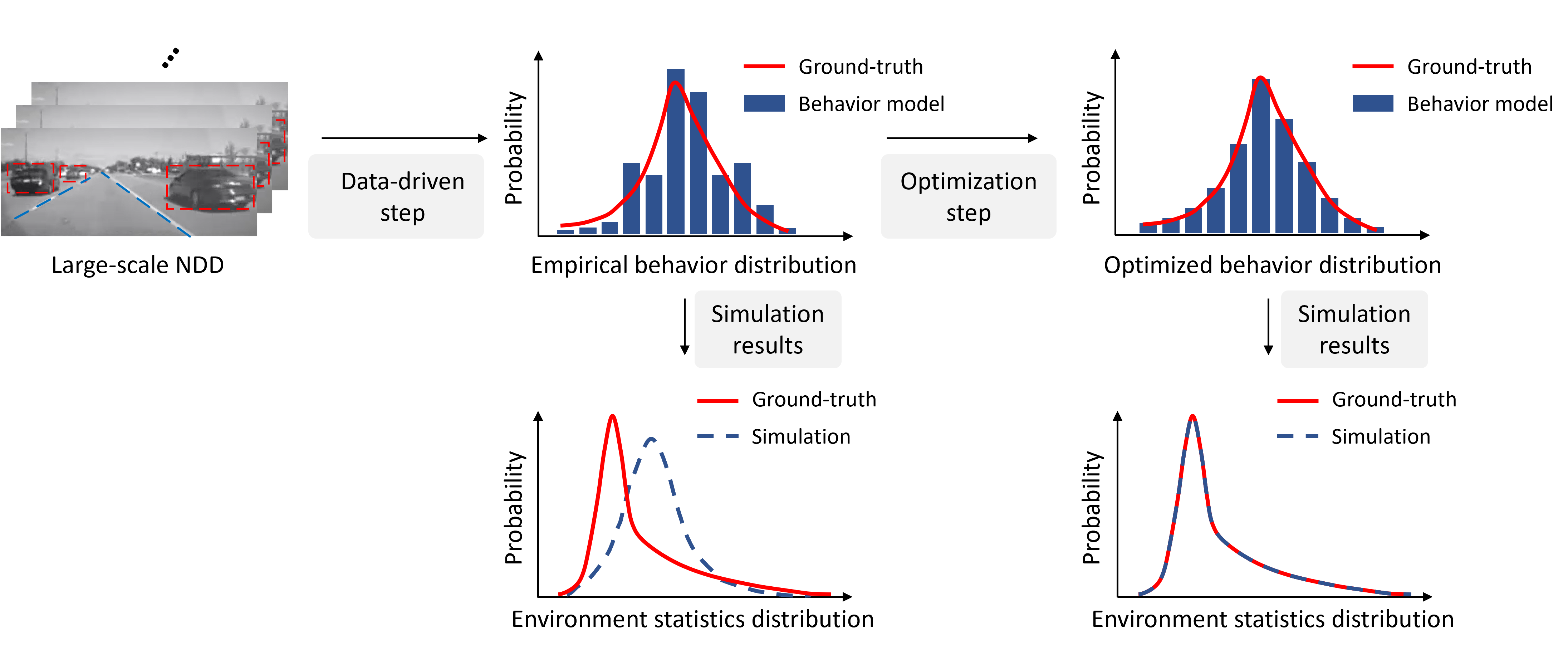}

\caption{Proposed framework pipeline. \label{fig:fig2-proposed-framework}}
\end{figure*}

To further illustrate the new requirements, we consider a simple car-following
scenario, where the leading vehicle is a human-driven vehicle and
the following vehicle is an AV, as shown in Fig. \ref{fig:mislead-AV-development}.
In the real-world driving environment, the behaviors of the leading
vehicle could have an underlying distribution, which leads to the
range distribution between two vehicles after one timestep (see Fig.
\ref{fig:mislead-AV-development}, middle, red curve). If the NDE
model cannot accurately represent the stochastic behaviors of the
leading vehicle, there could exist an inconsistency between the range
distributions in simulation (Fig. \ref{fig:mislead-AV-development},
middle, blue dashed line) and real-world. After several time steps,
this inconsistency will be accumulated and amplified, which could
lead to significant estimation errors of the accident rate (Fig. \ref{fig:mislead-AV-development},
right, yellow area). In this example, the accident rate will be underestimated in simulation,
which could mislead the further development and deployment of AVs. Therefore,
a realistic NDE with distributionally consistent environment statistics
is critical for AV testing, which is a challenging requirement. To the best of our knowledge, we are the first to identify and systematically investigate this new NDE modeling requirement, and there is no existing method that can fulfill
this requirement. 

To fill this research gap, we propose a data-driven optimization-based NDE modeling framework in this paper, which can ensure the distributional consistency of vehicle microscopic behaviors. The overall pipeline of the framework includes two major
steps as shown in Fig. \ref{fig:fig2-proposed-framework}. In the
first data-driven step, we propose to directly construct human driving
behavior models using empirical distributions from the large-scale
naturalistic driving data (NDD). Specifically, the driving behaviors are modeled by two longitudinal models (i.e., free driving and car following) and four lateral models considering
different driving situations, and each model is described as action distributions for different states. For example, the car-following behaviors can
be modeled by the acceleration distribution of the following vehicle
conditional on different self speeds, relative speeds, and relative
distances with the preceding vehicle. These empirical behavior models
serve as basic models. If the dataset is sufficiently accurate, diverse,
and large, the empirical distributions can accurately characterize human
driving behaviors in different situations. However, due to the limitation
of data quantity and unavoidable data noise, the obtained empirical
distributions may still have small inaccuracy compared with the ground truth. As illustrated in Fig. \ref{fig:mislead-AV-development}, the small inaccuracy can be accumulated and compounded with the simulation,
so the resultant simulation environment will deviate from the realistic
environment. 

To further tackle the error accumulation issue, the second step of the framework is to refine empirical behavior models to minimize the accumulated errors using optimization methods. To
achieve this goal, the key is to model the long-term effects of the error accumulation. In this study, by modeling the vehicle state evolution as a Markov chain, the long-term effects of the error accumulation could be modeled based on the stationary distribution of the Markov chain. Then, an optimization problem can be formulated to minimize the
accumulated errors by adjusting the empirical behavior models.  
In this way, the error accumulation could
be reduced, which results in the NDE model with more accurate distributions.
This provides opportunities to conduct high-fidelity long-time simulation
for full-length trip evaluation of AV\textquoteright s performance
\cite{waymo2020-Waymo-SimulationCity}. Using the large-scale real-world
NDD collected by the University of Michigan, Ann Arbor, we validate
the performance of the proposed method for a multilane highway driving
environment. Compared with existing models, the proposed method demonstrates
superior performance regarding the distributional accuracy. To further validate the capability for AV testing,
the generated NDE is also utilized to test the safety performance of an
AV agent.

In summary, the main contributions of this paper are threefold: first, the new requirements of NDE modeling are identified for AV testing purposes, which cannot be satisfied by most existing methods;
second, a novel modeling framework is proposed to generate the NDE that
is distributionally consistent with the real-world driving environment;
third, the proposed method is validated using large-scale real-world
NDD and the generated NDE is further validated by testing an AV model.

The rest of this paper is organized as follows. In Section \ref{sec:Related-Work},
we provide a brief review of existing NDE modeling methods. In Section \ref{sec:Data-driven-NDE-modeling},
we introduce the data-driven step to construct empirical behavior models using large-scale
real-world NDD and evaluate their performances in a multilane highway
driving environment simulation. In Section \ref{sec:Robust-modeling-framework},
to account for the error accumulation problem, the robust modeling step is proposed to optimize
the empirical behavior models to obtain a distributionally
consistent NDE. In Section \ref{sec:Performance-evaluation},
the performance of the proposed method is validated and compared with existing models. An AV testing experiment is further conducted
to demonstrate the capability of the proposed NDE in this section.
Finally, Section \ref{sec:Conclusions} concludes the paper and lays
out some future directions.

\section{Related Work\label{sec:Related-Work}}

In this section, we provide a brief overview of existing driving behavior
models, especially focusing on how the behavior stochasticity is modeled
and whether they can achieve distribution-level accuracy. A general
form of the existing driving behavior model can be expressed by
\begin{equation}
u\left(t\right)=\psi\left(S\left(t\right),\theta\left(t\right)\right)+\epsilon\left(t\right),
\end{equation}
where $t$ denotes the time, $u\left(t\right)$ denotes the action
(e.g., longitudinal acceleration) of the vehicle at the $t$-th time
instance, $S\left(t\right)$ denotes the states of the ego-vehicle
and surrounding vehicles that have the influence on the ego-vehicle's
decision making, $\psi\left(\cdot\right)$ denotes the model that
maps from the state space to the action space, $\theta\left(t\right)$
denotes model parameters that could be deterministic or stochastic,
and $\epsilon\left(t\right)$ denotes the additive noise term. 

Most traditional models \cite{wiedemann1974-Wiedman74,gipps1981,gipps1986-LC-model,treiber2000-IDM,kesting2007-MOBIL}
are deterministic and cannot capture the stochastic nature of human
driving behaviors. For these models, the model parameters $\theta$
remain fixed during the simulation and no external noise term $\epsilon$
is added. Although the parameters of these models can be calibrated
using real-world data \cite{punzo2012-CF-calibration,sangster2013-CF-calibration,li2016-CF-calibration,zhu2018-CF-calibration,osorio2019-CF-calibration},
they cannot capture the stochastic behaviors of human drivers. Recently,
increasing studies have used machine learning-based methods to fit
the behavior model $\psi$ using neural networks \cite{wang2017-deep-learning-CF,huang2018-deep-learning-CF,zhu2018-RL-CF,zhang2019-deep-learning-CF,xie2019-deep-learning-LC}.
By utilizing large-scale naturalistic driving datasets, these methods
aim to better reproduce the observed trajectories of human drivers.
However, the problem of lacking accurate stochasticity still remains unsolved.

The stochasticity can be incorporated into the model to a certain extent by introducing
external noise term $\epsilon\left(t\right)$ and/or stochasticity
to model parameters $\theta\left(t\right)$. For external noise $\epsilon\left(t\right)$,
the most commonly used one is the Gaussian noise \cite{laval2014parsimonious-stochastic-CF-Gaussian,he2015simple-stochastic-CF-Gaussian,treiber2017-stochastic-IDM,kuefler2017-GAIL-CF,bhattacharyya2020-GAIL-CF}.
However, the external addition of the Gaussian noise cannot realistically
depict human driving behaviors since the interaction in different
driving conditions is highly complex and does not always follow the
simple Gaussian distribution \cite{yeo2008-asymmetric,yang2010development-Errorable-CF-model,li2017-asymmetric}.
The inaccuracy may not affect significantly for the traffic flow analysis,
for example, these models can reproduce real-world traffic phenomena
like traffic oscillations and the fundamental diagram. However, it
will cause significant evaluation biasedness on AV performance. In a few studies \cite{yang2010development-Errorable-CF-model}, the log-normal and extreme value distributions are applied.
Although these distributions are more realistic than the Gaussian,
the accuracy is still limited due to the model flexibility for fitting
complicated driving behaviors in all situations using the same distribution.

\begin{figure*}[tp]
\centering
\subfloat[\label{fig:fig3a-6-categories}]{\includegraphics[width=0.9\textwidth]{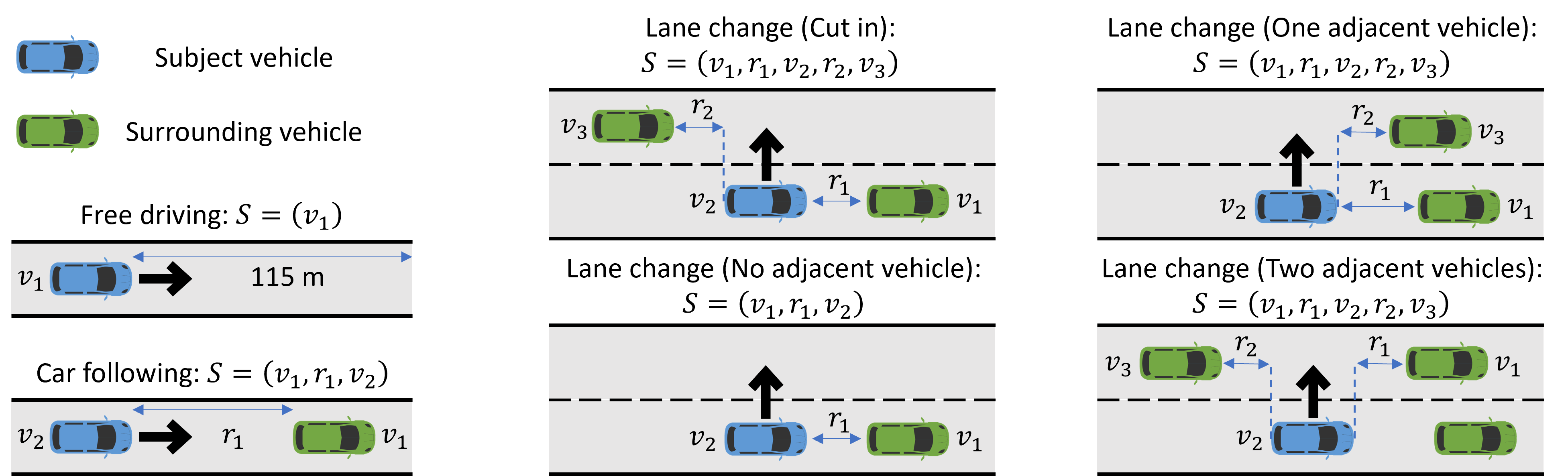}

}

\subfloat[\label{fig:fig3b-behavior-distribution-example}]{\includegraphics[width=0.9\textwidth]{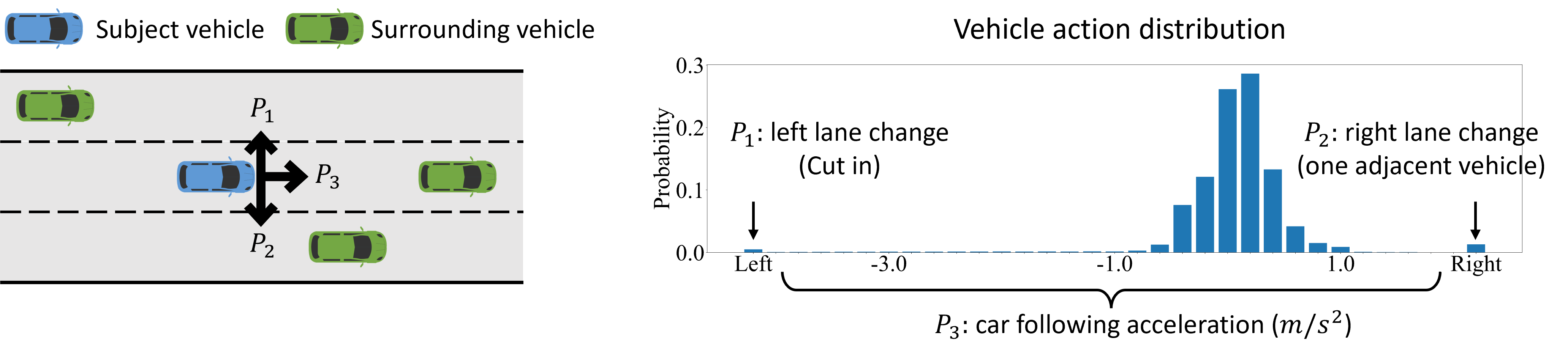}

}

\caption{Illustration of empirical behavior models: (a) longitudinal and lateral
behavior models in this study, (b) illustration example of how to simulate a vehicle's action at a decision moment.}
\end{figure*}

Besides adding an external noise $\epsilon\left(t\right)$, the stochasticity
can also be incorporated into model parameters $\theta\left(t\right)$.
For example, Treiber et al. \cite{treiber2017-stochastic-IDM}
proposed indifference regions in the form of action points to the
model, so the new acceleration will be executed only when the change
is greater than a threshold parameter, which follows a uniform distribution.
Yang et al. \cite{yang2010development-Errorable-CF-model} considered different
human errors during car-following situations, where the length of time delay
and distraction interval are modeled by exponential
distribution and lognormal distribution according to real-world data,
respectively. Game theory is applied in \cite{talebpour2015modeling-game-theory}
to model stochastic lane-change behavior. Hamdar et al. \cite{hamdar2008modeling-prospect-theory1,hamdar2015behavioral-prospect-theory2}
proposed a utility-based stochastic car-following model where acceleration
probability follows a continuous logit model. Again, these methods do not concern with the distributional accuracy
of the driving behaviors after the stochasticity is introduced. There were two notable exceptions
in \cite{wang2009markov-headway-distribution,chen2010-CF-model},
the authors proposed stochastic car-following models to capture the
distribution of time headway. They validate that the headway distribution
of the simulation environment is consistent with the real world. However,
only car-following behavior is considered in these studies, so they can only simulate
the single-lane road. More importantly, the error accumulation issue has not been considered, which could severely distort the distribution for long-term simulation of NDE. 
More related studies can be found in \cite{zheng2014-LC-review,saifuzzaman2014incorporating-CF-review,chao2020survey-Driving-models-review} and references therein.

Notwithstanding the related studies, most existing methods cannot
capture the accurate distribution of stochastic human driving behaviors.
Therefore, the driving environment that is generated by these behavior
models cannot accurately reproduce simulation environment statistics,
which is critical for AV testing.

\section{Data-driven NDE modeling\label{sec:Data-driven-NDE-modeling}}

In this section, we propose a simple yet effective data-driven method for NDE modeling leveraging large-scale NDD. Specifically, six empirical behavior models are constructed  including free-driving, car-following, and four lane-changing behaviors with different driving conditions, and then the NDE can be generated by combining the six empirical behavior models according to the driving condition at each time step (Section \ref{subsec:Empirical-behavior-models}). To construct each empirical behavior model, the large-scale NDD is processed and utilized in Section \ref{subsec:Naturalistic-driving-dataset-processing}. Then, a multi-lane highway driving environment is simulated to evaluate the performance of the empirical behavior models in Section \ref{subsec:before-optim-performance}, which validates the data-driven method and further motivates the robust modeling step in Section \ref{sec:Robust-modeling-framework}.

\subsection{Empirical behavior models\label{subsec:Empirical-behavior-models}}

\begin{figure*}[t]
\begin{centering}
\includegraphics[width=1\textwidth]{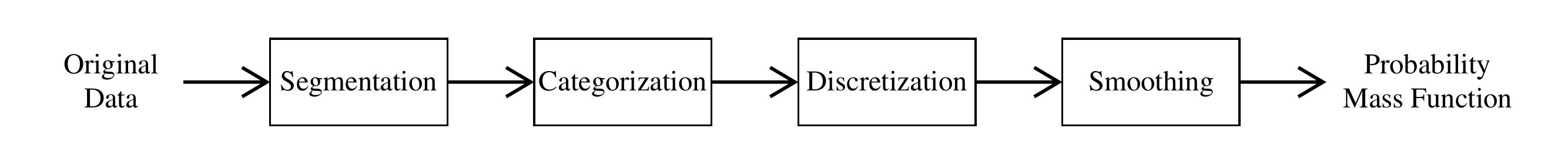}
\par\end{centering}
\caption{Data processing flow chart. \label{fig:fig4-Data-processing-flow}}
\end{figure*}

To construct the NDE, both longitudinal and lateral behaviors of human drivers need to be modeled based on the vehicle's own state and its surrounding situations $S$. For example, human drivers need to decide their longitudinal accelerations and decide whether to take a lateral lane-change maneuver. In this study, six behavior models are proposed including free-driving, car-following, and four lane-changing behaviors with different driving conditions, as shown in Fig. \ref{fig:fig3a-6-categories}. Specifically, the vehicle acceleration in the free-driving case is modeled that depends only on its current velocity, while the acceleration in the car-following case is modeled that depends on the velocity, range (relative position), and range rate (relative speed) of the subject vehicle and its preceding vehicle. To capture the lane-changing probability in different conditions, four lane-changing models are proposed, which output the lane change probability of the subject vehicle at each moment. For example, in the cut-in lane change situation, the lane-changing probability depends on velocities and distances between the subject vehicle and the preceding vehicle in the current lane and the vehicle behind in the target lane. We note that more lane-changing models could be constructed in this framework by dividing the driving conditions into more categories if needed. 

After constructing the six behavior models, the NDE can be generated by combining the behavior models according to the driving condition at each time step. Taking Fig. \ref{fig:fig3b-behavior-distribution-example} (left) as an example, the subject vehicle can take left lane change, keep car-following, or take right lane change at the moment. The left lane change behavior can be categorized as a cut-in behavior, where the lane-changing probability $P_{1}$ can be obtained from the cut-in behavior model. The right lane change behavior can be categorized as a lane change with one adjacent vehicle, where the lane-changing probability $P_{2}$ can also be obtained by the corresponding behavior model. Moreover, the
longitudinal acceleration probability $P_{3}$ can be obtained by the car-following behavior model. After normalization, we can obtain the action distribution of the vehicle as shown in Fig. \ref{fig:fig3b-behavior-distribution-example} (right). Then, the subject vehicle's action will be sampled from this distribution and used to update its state to the next time step. To simplify the modeling process, longitudinal acceleration is assumed zero if the vehicle is making a lane change behavior. Also, if there is no vehicle in front, the ego-vehicle will not take lane-changing behavior. By repeating this process for all vehicles and time steps, the NDE can be generated.

The remaining question is how to construct the six behavior models with distributional accuracy. In this study, we propose to directly estimate empirical behavior distributions as the behavior models by leveraging large-scale NDD. As the NDD records all the information needed for the human driving behaviors, accurate empirical behavior models could be constructed if using sufficient amount of data with perfect quality. Although the 
actual data is usually limited by the data quality and quantity, these empirical behavior models could provide a good foundation and can be further improved as discussed in Section \ref{sec:Robust-modeling-framework}. Specifically, for each behavior model, we obtain the empirical probability $P\left(a|S\right)$ for all the vehicle
actions $a\in\mathcal{A}$ at all discretized states $S\in\mathcal{S}$,
where $\mathcal{A}$ denotes the action space, and $\mathcal{S}$ denotes the state space. Let $F\left(S\right)=\left[P\left(a_{1}\right|S),\cdots F\left(a_{\left|\mathcal{A}\right|}|S\right)\right]$ denote the probability mass function
under a certain state $S$, then the empirical behavior model can be denoted by
\begin{equation}
F=\left[F\left(S_{1}\right),\cdots F\left(S_{\left|\mathcal{S}\right|}\right)\right] \in\mathbb{R}^{\left|\mathcal{S}\right|\times\left|\mathcal{A}\right|}.
\label{eq:behavior-model}
\end{equation}
Next, we will introduce how to process the NDD and construct $F$ for all the six behavior models.

\subsection{Naturalistic driving dataset processing \label{subsec:Naturalistic-driving-dataset-processing}}

To construct empirical behavior models, we utilized large-scale NDD
from the Integrated Vehicle Based Safety System (IVBSS) dataset \cite{sayer2011-IVBSS}
and the Safety Pilot Model Deployment (SPMD) dataset \cite{bezzina2014-SPMD}
at the University of Michigan Transportation Research Institute (UMTRI).
In the IVBSS program, 108 drivers ranging from 20 to 70 years old
were recruited. Each participant drove the IVBSS vehicle equipped
with the data acquisition system (DAS) for 6 weeks. The relative distance
and speed with the leading vehicle are recorded by radar at 10 Hz.
The SPMD program covered over 34.9 million travel miles and included
98 vehicles equipped with the DAS and Mobileye
to record human naturalistic driving behaviors. The data were also
recorded at 10 Hz with positions, speeds, and accelerations of ego-vehicles,
relative speeds with surrounding vehicles, and both longitudinal and
lateral distances between vehicles and lane markings. We queried partial
datasets with the following criteria: (1) vehicle was traveling at
a speed between 20 m/s and 40 m/s; (2) dry surface condition; (3)
daylight condition. The resulting dataset includes approximately 8,200 driving hours data.

The data processing consists of four steps  including segmentation, categorization, discretization, and smoothing, as shown in Fig. \ref{fig:fig4-Data-processing-flow}.
Specifically, the original data were first segmented into trajectories and then
categorized into specific groups based on the six driving situations
defined in empirical behavior models. Then, a smoothing technique
was applied to the discretized action distribution and finally, we
could obtain the probability mass functions for each group, which
constituted the six empirical behavior models.  More details of the data processing steps can be found in Appendix. \ref{sec:Appendix-NDD-processing}. Fig. \ref{fig:fig6} demonstrates examples of constructed empirical behavior models. Specifically, Fig. \ref{fig:fig6a-FF} and Fig. \ref{fig:fig6b-CF}
show examples of the vehicle longitudinal acceleration distributions
in free-driving and car-following situations, respectively. For the
car-following case, Fig. \ref{fig:fig6b-CF} indicates acceleration
distribution when the ego-vehicle and its preceding vehicle have the
same speed and their range is 30 meters. We can find that for both
free-driving and car-following cases, the mean of acceleration is
around zero, which is consistent with the intuition. Compared with
the car-following situation, the probability of acceleration greater
than zero is generally higher in the free-driving case, which is reasonable
as well.

\begin{figure}
\centering
\subfloat[\label{fig:fig6a-FF}]{\begin{centering}
\includegraphics[width=0.8\columnwidth]{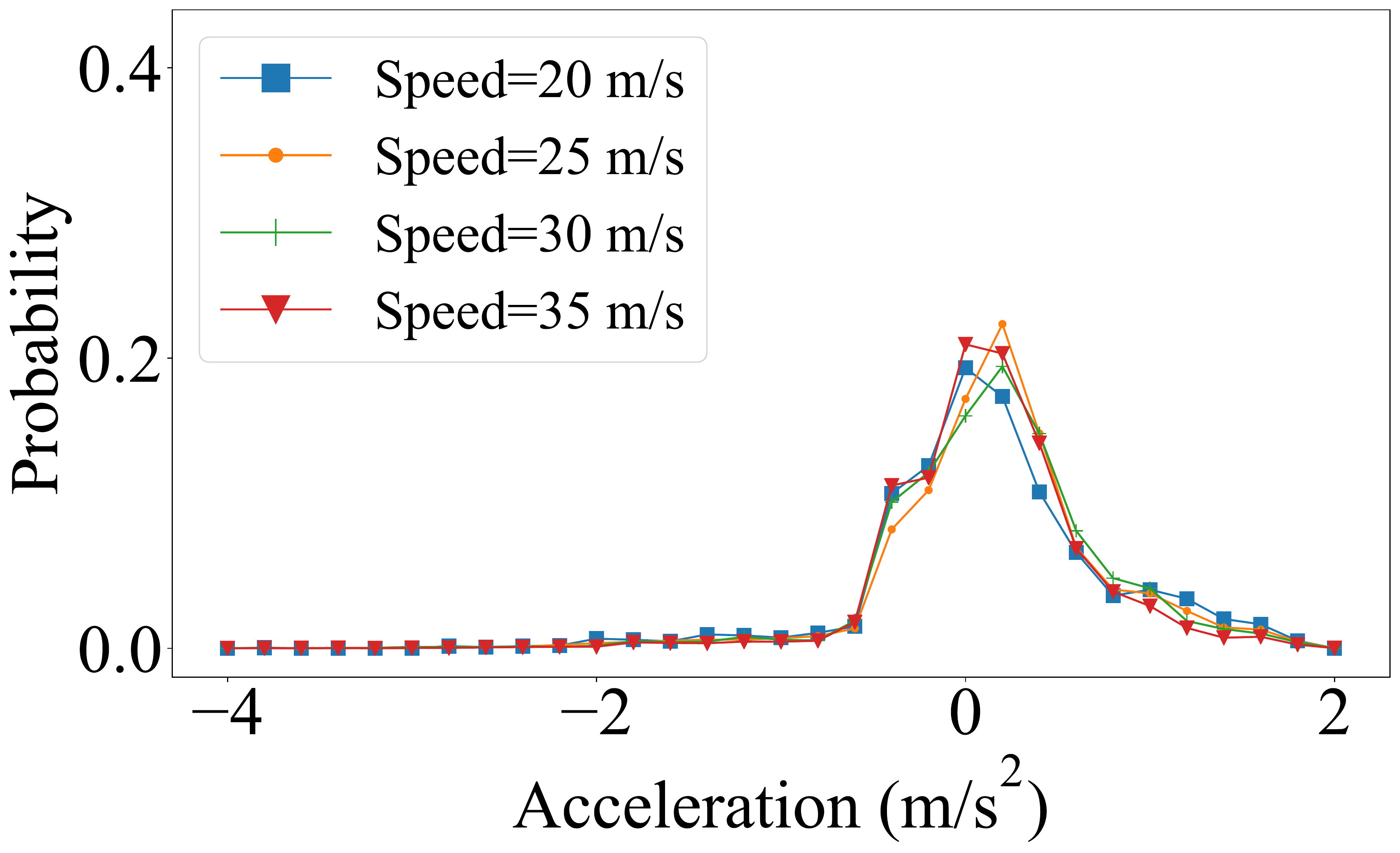}
\par\end{centering}
}

\subfloat[\label{fig:fig6b-CF}]{\begin{centering}
\includegraphics[width=0.8\columnwidth]{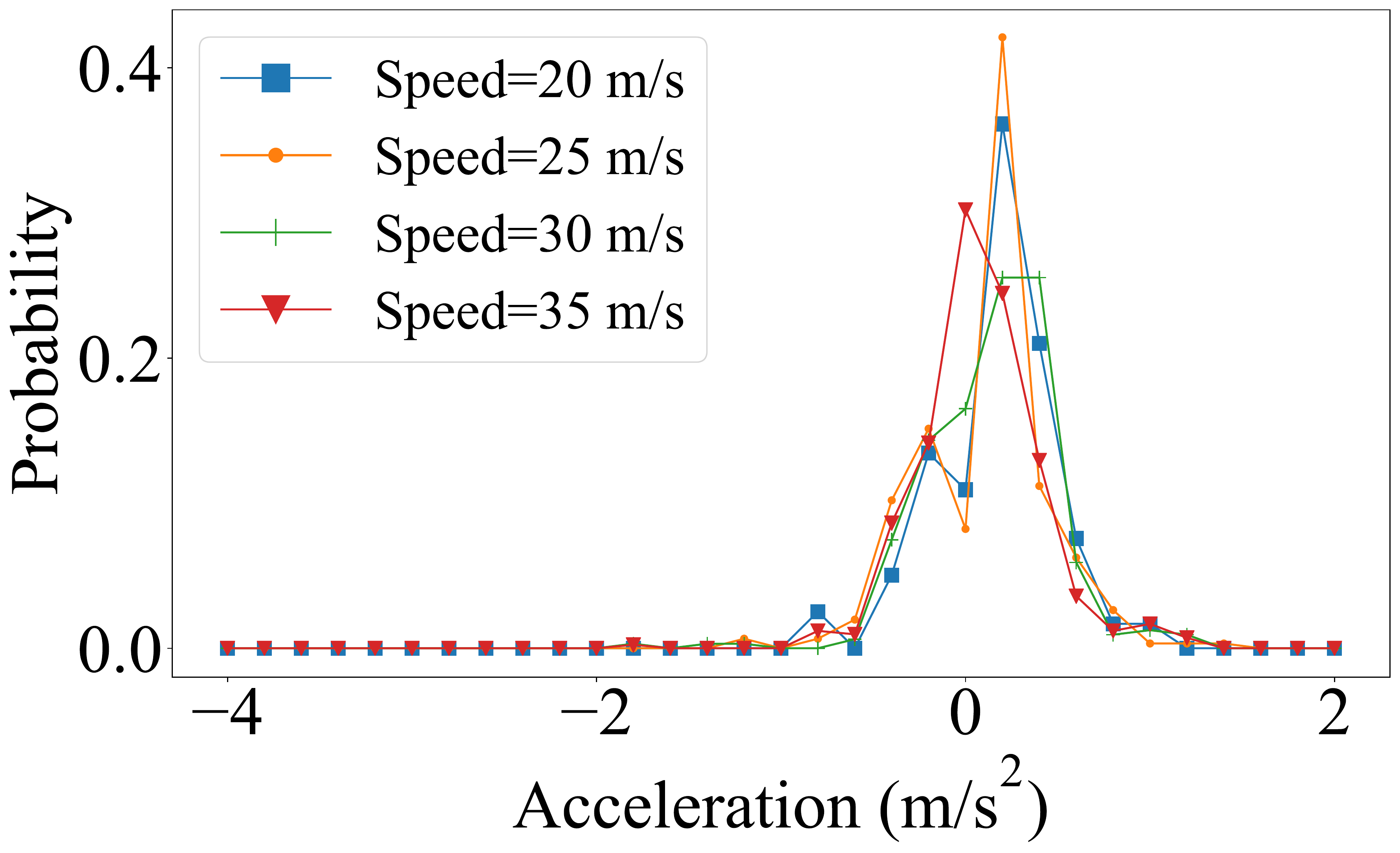}
\par\end{centering}
}\caption{Examples of empirical behavior models: (a) Free driving, (b) Car following
($r_{1}=30\ m,v_{1}=v_{2}$). \label{fig:fig6}}

\end{figure}

\subsection{Performance evaluation of empirical behavior models \label{subsec:before-optim-performance}}
\begin{figure}
\begin{centering}
\includegraphics[width=0.8\columnwidth]{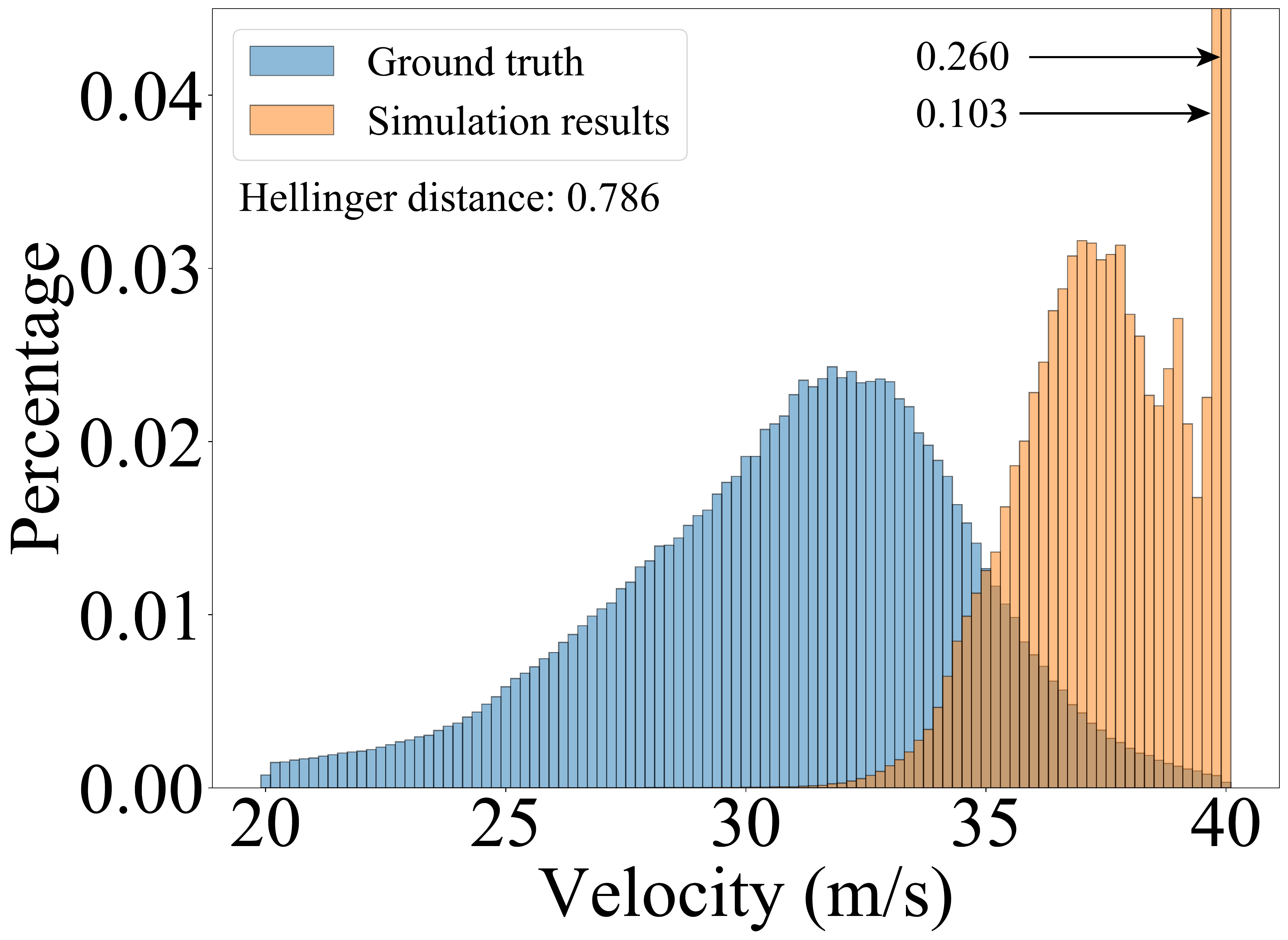}
\par\end{centering}
\begin{centering}
\includegraphics[width=0.8\columnwidth]{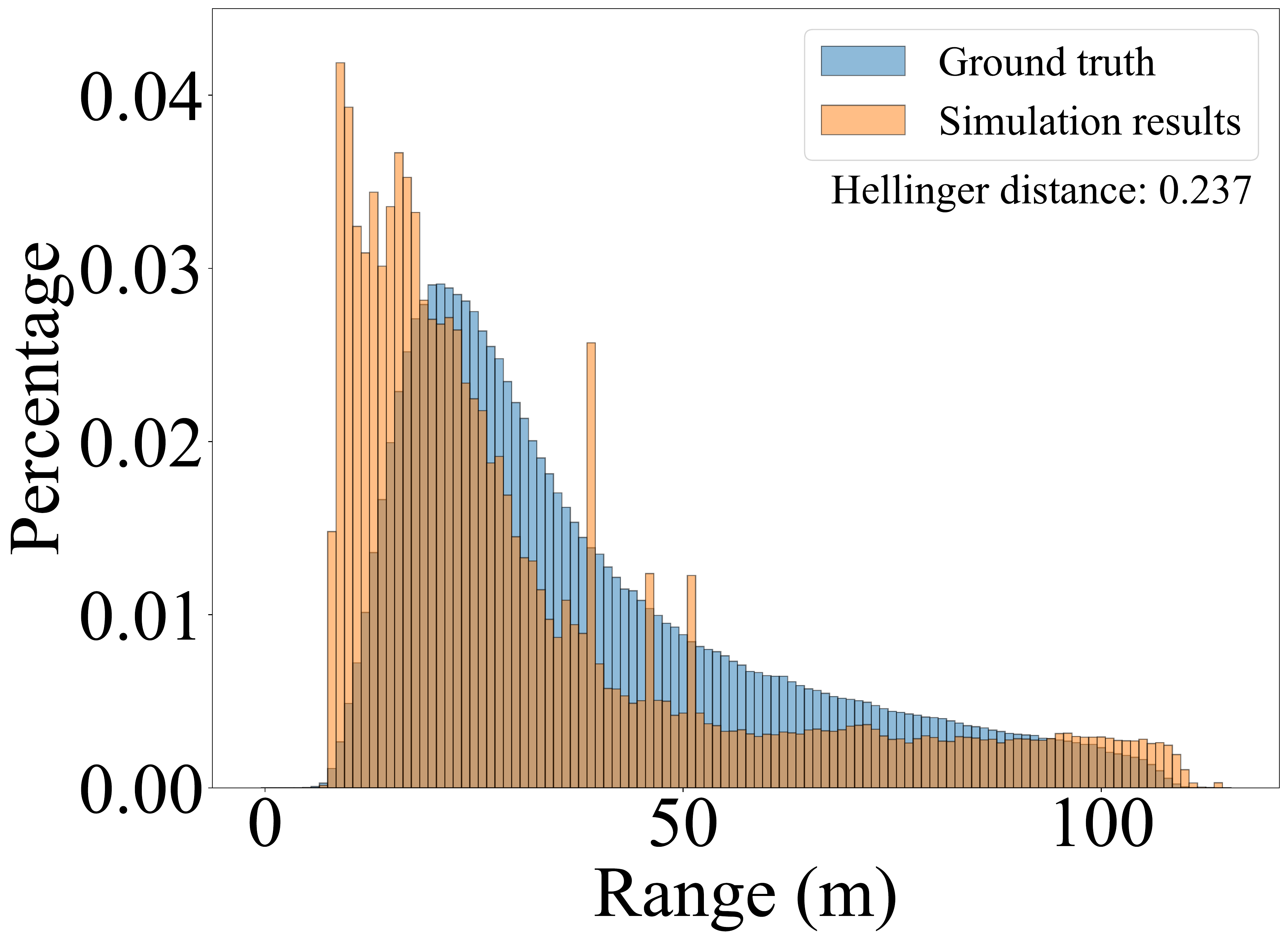}
\par\end{centering}
\caption{Velocity and range distributions of the NDE using empirical behavior
models. \label{fig:fig7-error-accum}}
\end{figure}

In this subsection, the performance of the NDE constructed by the six empirical behavior models is evaluated in a three-lane highway simulation, as illustrated in Fig. \ref{fig:fig3b-behavior-distribution-example} (left). 
The detailed simulation settings can be found in Appendix. \ref{sec:AppendixB-sim-settings}. The Hellinger
distance \cite{Hellinger-distance} is used to quantitatively measure the dissimilarity between
the simulated distribution and the true distribution. The Hellinger
distance ranges from 0 to 1, and the smaller the measurement, the
better the model performance.
To demonstrate the performance, the background vehicles velocity and range distributions, which are important for the AV testing, are investigated as shown in Fig.
\ref{fig:fig7-error-accum}. Results show that although the distributions can roughly capture the trends of the real-world distributions, there still exist significant distributional inconsistency, particularly for the vehicle velocity. This inconsistency is caused by the error accumulation of the empirical behavior models, where the small model errors caused by the limited data quality and quantity are accumulated and amplified along with the simulation steps, as illustrated in Fig. \ref{fig:mislead-AV-development}.
To address this
issue, the robust NDE modeling step is developed in the next
section to further improve the empirical models.

\section{Robust NDE modeling\label{sec:Robust-modeling-framework}}

In this section, the robust NDE modeling step is proposed to refine empirical behavior models to minimize the accumulated errors using optimization methods. To achieve this goal, the key is to model the long-term effects of the error accumulation. Specifically, by modeling the vehicle state evolution as a Markov chain, the long-term effects of the error accumulation can be characterized by the stationary distribution of the Markov chain. Then, an optimization problem is formulated to minimize the
accumulated errors by adjusting the empirical behavior models, which results in the NDE model with more accurate distributions.
In the following paragraphs, we first propose the optimization framework in 
Section \ref{subsec:Overall-optimization-framework} and then apply the framework on the longitudinal behavior models in 
Section \ref{subsec:Longitudinal-behavior-model}.


\subsection{Optimization framework \label{subsec:Overall-optimization-framework}}

In order to solve the error accumulation problem, we need to measure the accumulated error
of the NDE generated by the empirical behavior models. One possible way is to simulate the
NDE, collect the data, and obtain the simulated NDE distribution. However, the computational
burden of this method is very heavy since a large number of simulations are needed to obtain an
accurate estimation. To overcome this problem, we propose to measure the
NDE distribution by analyzing the stationary distribution of the NDE Markov chain. By using
this analytical method, the NDE stationary distribution serves as an accurate approximation of
the simulated environment, which reflects the performance of the empirical behavior models. By fitting the simulated stationary distribution with real-world ground-truth, we can improve the NDE accuracy and solve the error accumulation issue.

\begin{figure}[tbh]
\begin{centering}
\includegraphics[width=1\columnwidth]{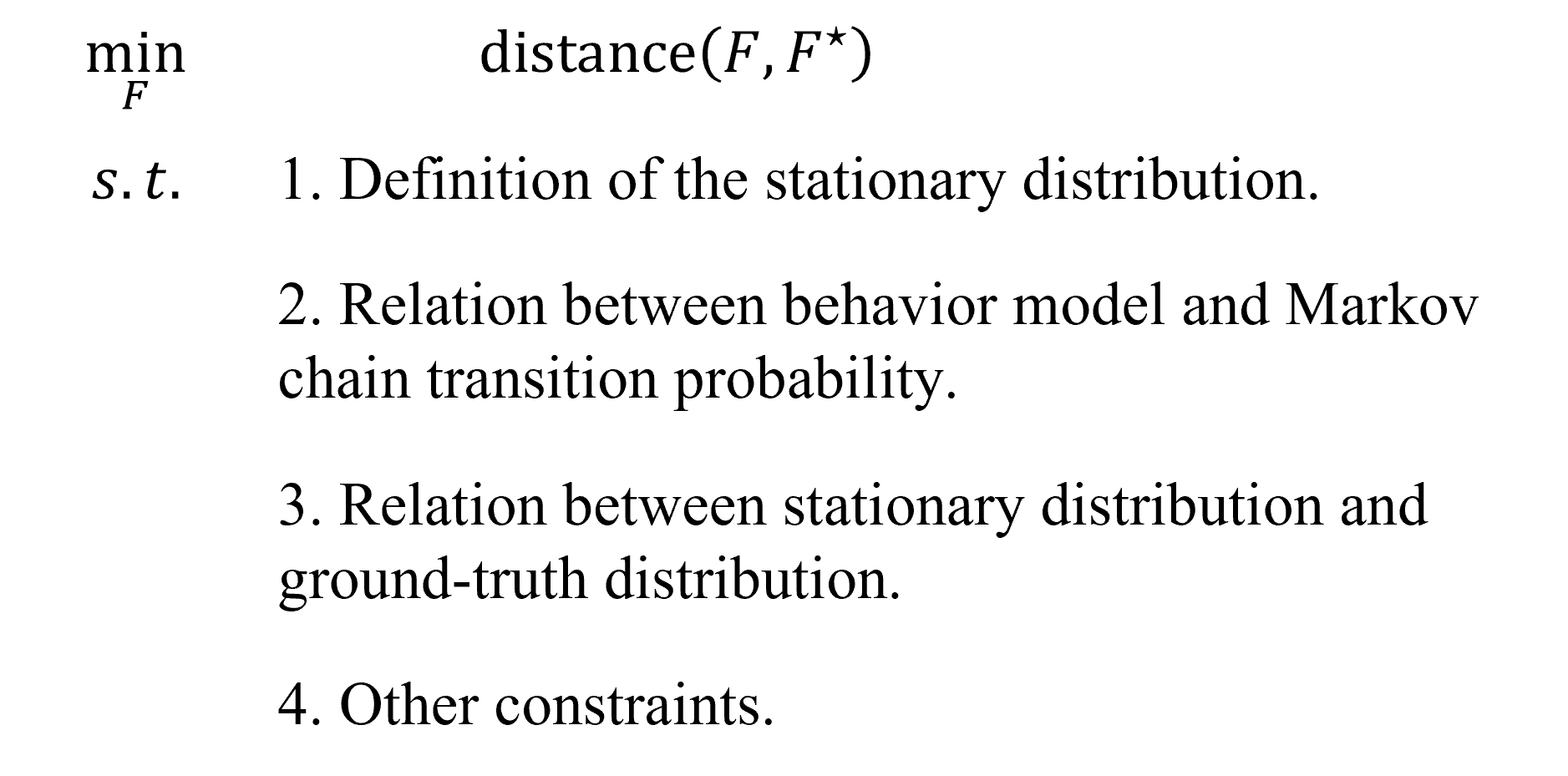}
\par\end{centering}
\caption{Overall formulation of the proposed robust modeling step.\label{fig:fig8-Overall-formulation-of}}

\end{figure}

Following this idea, the optimization framework can be formulated as in Fig. \ref{fig:fig8-Overall-formulation-of}.
The decision variable is the vehicle behavior model $F$ as shown in Eq. (\ref{eq:behavior-model}), and the
objective is to minimize the adjustment to the empirical
behavior model $F^{\star}$ while ensuring the accuracy of the stationary distribution. To achieve this objective, there are generally four sets of constraints
in the optimization formulation. The first set of constraints is the standard definition of stationary distribution, which indicate that the state
will always follow its stationary distribution after reaching the
steady state. The second set of constraints describes the relationship
between the behavior model and Markov chain state transition probability.
The stochastic vehicle behavior model outputs actions for the next
time step and therefore determines the state transition process. The
third set of constraints is to match the stationary distribution
of the simulation with the real-world ground-truth distribution, which is the key to reducing the accumulated errors.
As a result, the simulated environment can be
guaranteed to fit the desired real-world statistics (e.g., velocity and range distributions) even after a long
simulation time horizon. The last set of constraints denotes other
standard requirements, such as non-negative constraints of probability
mass functions, normalization of stationary distribution, etc.

\begin{figure*}
\begin{centering}
\subfloat[\label{fig:fig9a-ours}]{\begin{centering}
\includegraphics[width=0.4\textwidth]{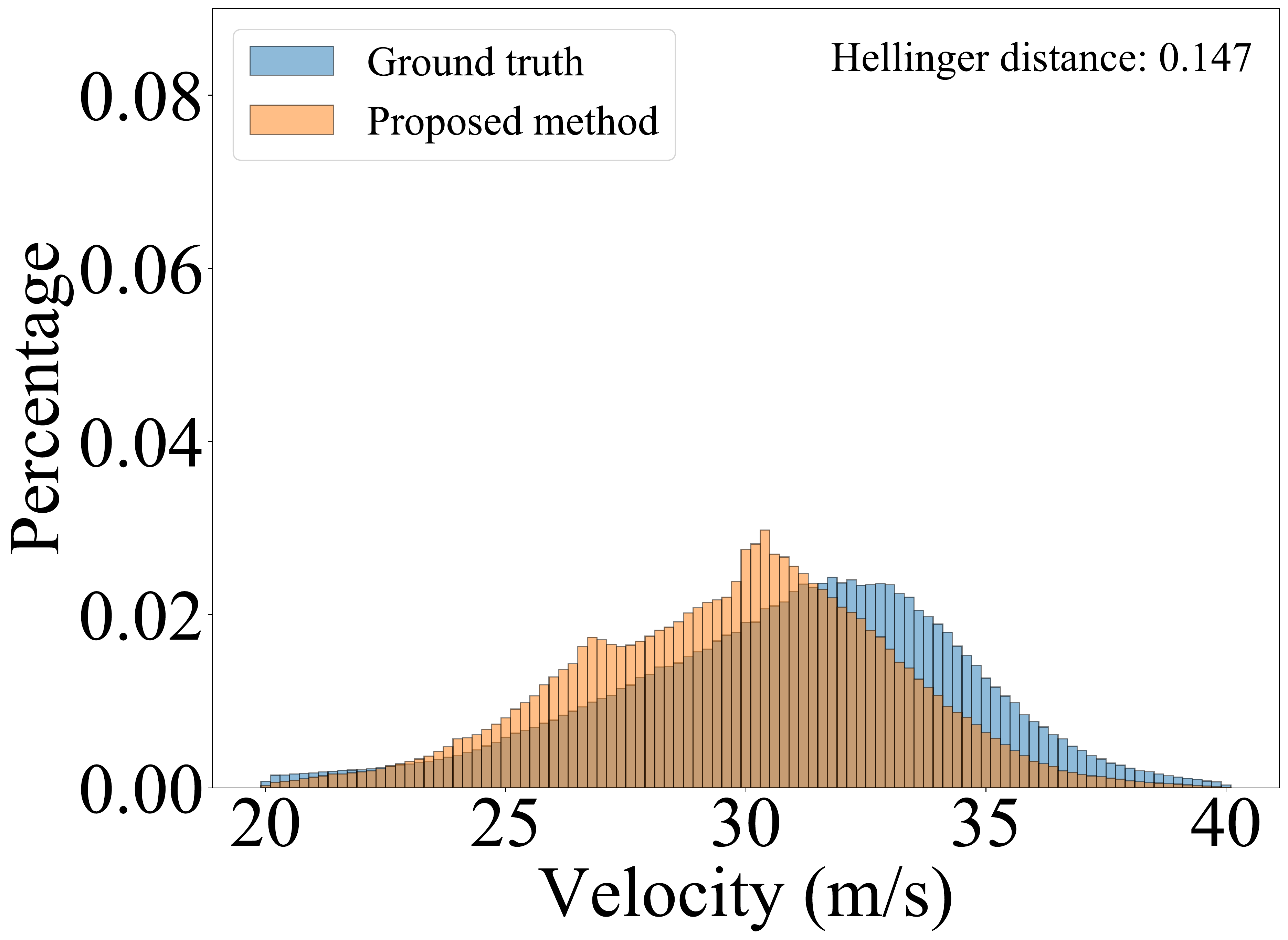}\includegraphics[width=0.4\textwidth]{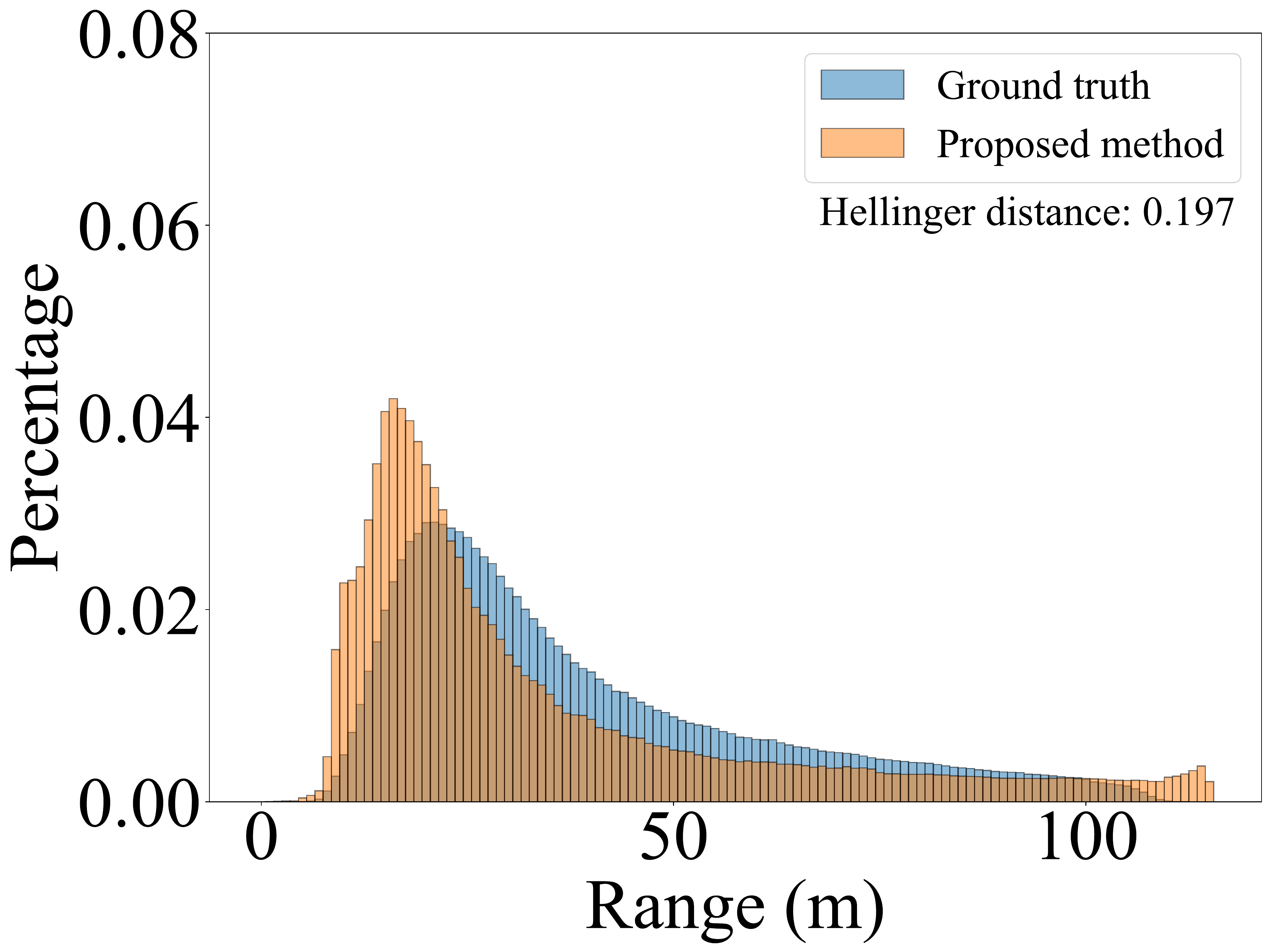}
\par\end{centering}
}
\par\end{centering}
\centering{}\subfloat[\label{fig:fig9b-sumo}]{\begin{centering}
\includegraphics[width=0.4\textwidth]{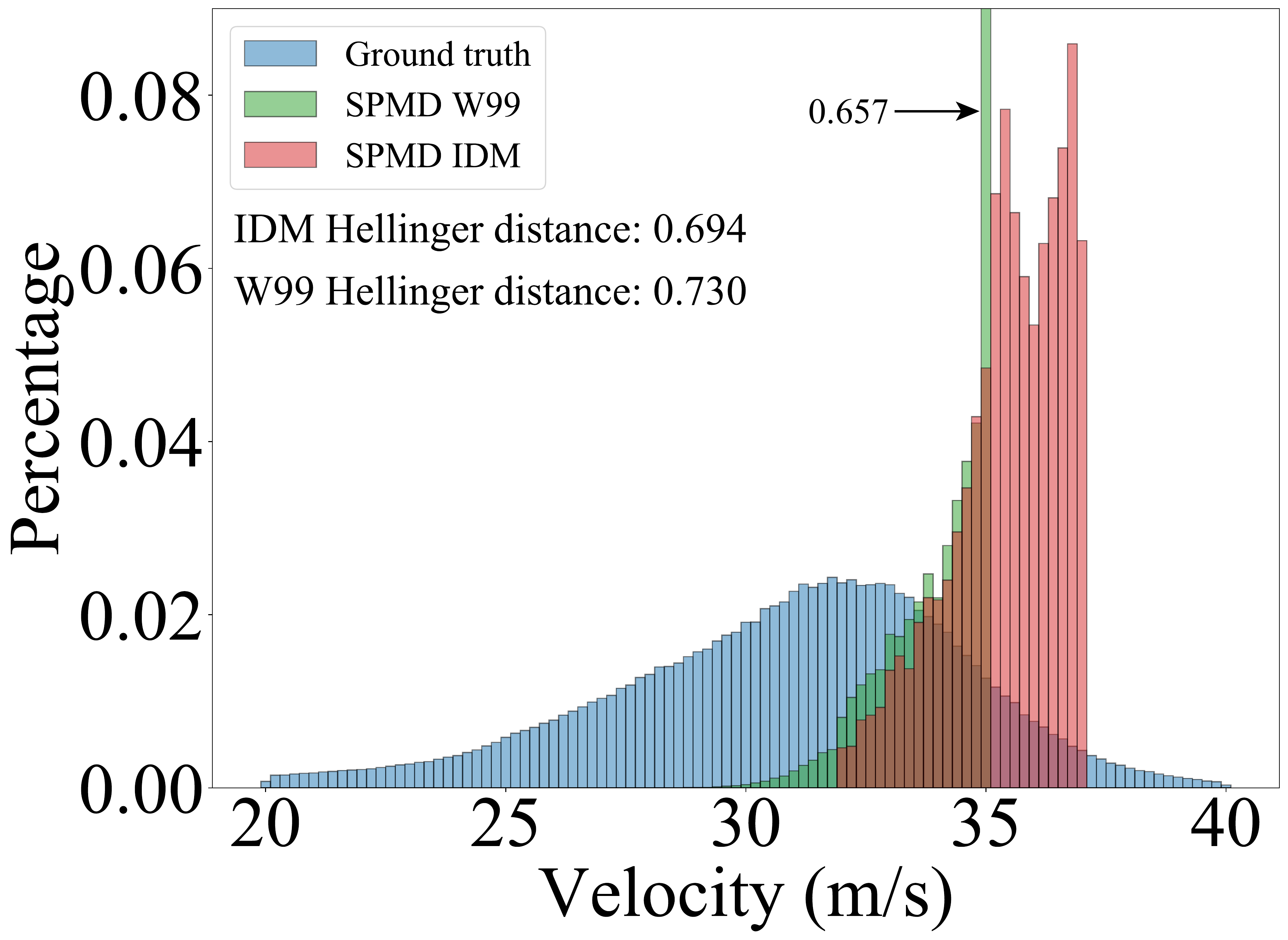}\includegraphics[width=0.4\textwidth]{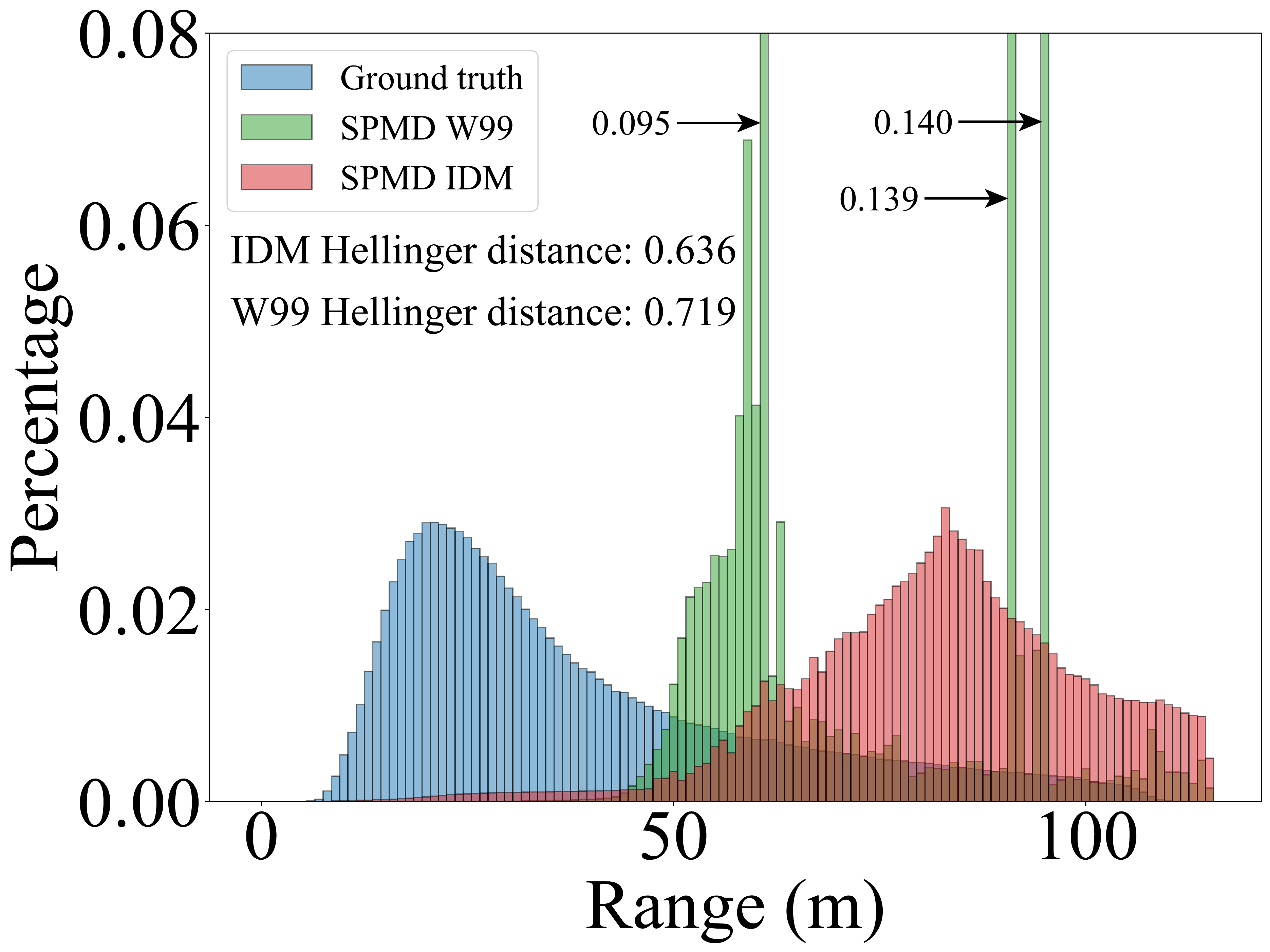}
\par\end{centering}

}\caption{Velocity and range distributions of the: (a) proposed method; (b)
SUMO simulator. \label{fig:fig9-main-results}}
\end{figure*}

\subsection{Optimization of longitudinal behavior models \label{subsec:Longitudinal-behavior-model}}

In this section, we apply the proposed framework to optimize the two empirical
longitudinal behavior models as a proof of concept, while keeping the four
empirical lateral behavior models unchanged. As the velocities cannot be well modeled by the empirical models as shown in Fig. \ref{fig:fig7-error-accum}, we choose the velocity distribution as the optimization target. 

For the free-driving behavior model, we define the discretized speed as the state of the Markov chain. It is easy to find that the Markov chain is finite, irreducible, and aperiodic, so there exists a unique positive stationary distribution $\pi$ \cite{grimmett2020-MC-reference}
satisfying

\begin{align}
\pi^{T}\textbf{P} & =\pi^{T},\label{eq:stationary-distribution-constraint}\\
\sum_{S\in\mathcal{S}}\pi_{S} & =1,\\
\pi & \succcurlyeq0,
\end{align}
where $\textbf{P}$ is the state transition probability matrix.
As the vehicle longitudinal acceleration depends only on its current speed in the free-driving situation, the state transition probability matrix $\textbf{P}$ is essentially a function of the behavior model $F$ in Eq. (\ref{eq:behavior-model}) as

\begin{equation}
\textbf{P}\left(S_{i},S_{j}\right)=G\left(F\right),\forall S_{i},S_{j}\in\mathcal{S},\label{eq:G(F)=00003DP}
\end{equation}
where $G\left(\cdot\right)$ is a linear mapping from the
longitudinal acceleration to the state transition. For example, if the
current speed falls in the state $S_{i}$, the next speed after the transition
is $S_{j}$, the time resolution is $\Delta t$, and the probability
of taking acceleration $a$ that satisfies the $S_{j}=S_{i}+a\cdot\Delta t$
is $p\left(a|S_{i}\right)$, then $\textbf{P}\left(S_{i},S_{j}\right)=p\left(a|S_{i}\right)$. 
Moreover, as the goal of the optimization is to match the vehicle stationary speed
distribution with the real-world speed distribution in the free-driving
situation, we have
\begin{equation}
\pi=\pi^{\star},
\end{equation}
where $\pi^{\star}$ is the ground-truth of the speed distribution in free-driving situations that is obtained from the large-scale NDD.

Finally, the optimization problem can be formulated as below:

\begin{eqnarray}
\min_{F} &  & \left\Vert F-F^{\star}\right\Vert _{Frob}\label{eq:optim-start}\\
\text{s.t.} &  & \pi^{T}\textbf{P}=\pi^{T},\label{eq:pi@P=00003Dpi-Summarized}\\
 &  & G\left(F\right)=\textbf{P}\left(S_{i},S_{j}\right),\forall S_{i},S_{j}\in\mathcal{S},\label{eq:G(F)=00003DP-Summarized}\\
 &  & \pi=\pi^{\star},\label{eq:consistency-Summarized}\\
 &  & \sum_{a\text{\ensuremath{\in}\ensuremath{\mathcal{A}}}}F\left(a|S\right)=1,\forall S\in\mathcal{S},\label{eq:normalization-start}\\
 &  & \sum_{S_{j}\text{\ensuremath{\in\mathcal{S}}}}\textbf{P}\left(S_{i},S_{j}\right)=1,\forall S_{i}\in\mathcal{S},\\
 &  & \sum_{S\in\mathcal{S}}\pi_{S}=1,\label{eq:normalization-end}\\
 &  & F,\textbf{P},\pi\succcurlyeq0.\label{eq:non-negativity}
\end{eqnarray}
The Frobenius norm $\left\Vert \cdot\right\Vert _{Frob}$ is adopted
to measure the distance between the optimized free-driving model $F$
and empirical free-driving model $F^{\star}$.
Comparing with the constraints discussed in Fig. \ref{fig:fig8-Overall-formulation-of}, 
Eq. (\ref{eq:pi@P=00003Dpi-Summarized}) denotes the definition of the stationary distribution, 
Eq. (\ref{eq:G(F)=00003DP-Summarized}) denotes the relation between behavior model and transition probability, 
Eq. (\ref{eq:consistency-Summarized}) denotes the distributional consistency between the stationary distribution and the ground-truth, and Eqs. (\ref{eq:normalization-start}-\ref{eq:non-negativity}) denote other constraints including the normalization requirements for the acceleration probability mass function,
state transition probability matrix, and stationary distribution,
respectively, and the non-negative
requirements. 
It can be found out that this is a linear programming problem that can be solved
efficiently using commercial solvers, for example, Gurobi \cite{gurobi}. 

For the car-following situation, the vehicle state is composed of
the speed of the subject vehicle $(v)$, range $(r)$, and range rate
$(rr)$ with the preceding vehicle. The state transition in the car-following
situation depends not only on the subject vehicle action but also
preceding vehicle action, which makes the optimization problem more complex. To solve this issue, we optimize the steady-state situation of the car-following model, which is a necessary condition regardless of the evolving process to the stationary distribution. As the preceding vehicle has 
reached the steady-state, the ego-vehicle state transition relies only upon its own action.
Then, the optimization problem can be formulated as the same as 
Eqs. (\ref{eq:optim-start}-\ref{eq:non-negativity}), where decision variables are the probability mass functions of the 
car-following accelerations ($F$ in Eq. (\ref{eq:behavior-model})), and Eq. (\ref{eq:pi@P=00003Dpi-Summarized})
is a three-dimensional joint
state distribution (i.e., $v$, $r$, and $rr$). It is also a linear programming problem that can be solved
efficiently.

\section{Performance evaluation\label{sec:Performance-evaluation}}
In this section, the performance of the proposed NDE
modeling framework is evaluated.

\subsection{Distributional consistency}
As the distributional consistency is critical for the generated NDE,  we first evaluate whether the proposed NDE can generate accurate velocity and range distributions, compared with the existing NDE baseline (i.e., SUMO \cite{lopez2018-SUMO}) and the empirical models constructed in Section \ref{sec:Data-driven-NDE-modeling}. Specifically, two
existing car-following models, the IDM \cite{treiber2000-IDM} and
the Wiedemann 99 model \cite{wiedemann99-W99}, and SUMO LC2013 lane-changing model \cite{erdmann2015-SUMO-LC-model} are selected as the existing NDE baseline,  which are widely applied in existing traffic simulators. For fair comparisons, the model parameters are calibrated with the SPMD NDD described in Section \ref{subsec:Naturalistic-driving-dataset-processing} using the calibration method developed in \cite{hammit2018case-Our-calibration,hammit2018case-code-Our-calibration}. More simulation settings can be found in Appendix. \ref{sec:AppendixB-sim-settings}.

Fig. \ref{fig:fig9-main-results} shows the results of the proposed NDE model and existing NDE models. It can be found that 
the proposed NDE model can significantly better reproduce the 
real-world velocity and range distributions than existing ones. 
Specifically, both the IDM and Wiedemann 99 models are concentrated 
in a small interval of velocity and range, while the real-world distributions range 
among a much wider interval. It is reasonable as these existing models are designed 
for accident-free purposes and therefore might be more conservative and deterministic. 
Besides using model parameters calibrated in this study, we also examine the performance 
using model parameters provided from the literature, where similar results are obtained (see Appendix. \ref{sec:AppendixE-literature-param} for more details). 

Compared with the results
before the optimization as shown in Fig. \ref{fig:fig7-error-accum}, the accuracy of the
simulated velocity distribution is also significantly improved, which validates the
effectiveness of the robust modeling step. To further quantify the performance improvement, the
Hellinger distances of all these models are calculated as listed in 
Table. \ref{tab:Quantitative-performance}. Results show that the empirical behavior 
models can achieve a better performance than the existing NDE models (similar performance in
velocity but significantly better performance in range), and the optimized models can
achieve the best performance. These results validate the effectiveness of both the data-driven and optimization steps and 
show that the proposed method can generate the NDE consistent 
with the real-world driving environment.

\begin{table}
\caption{Quantitative performance evaluation for different methods.\label{tab:Quantitative-performance}}
\begin{centering}
\includegraphics[width=0.75\columnwidth]{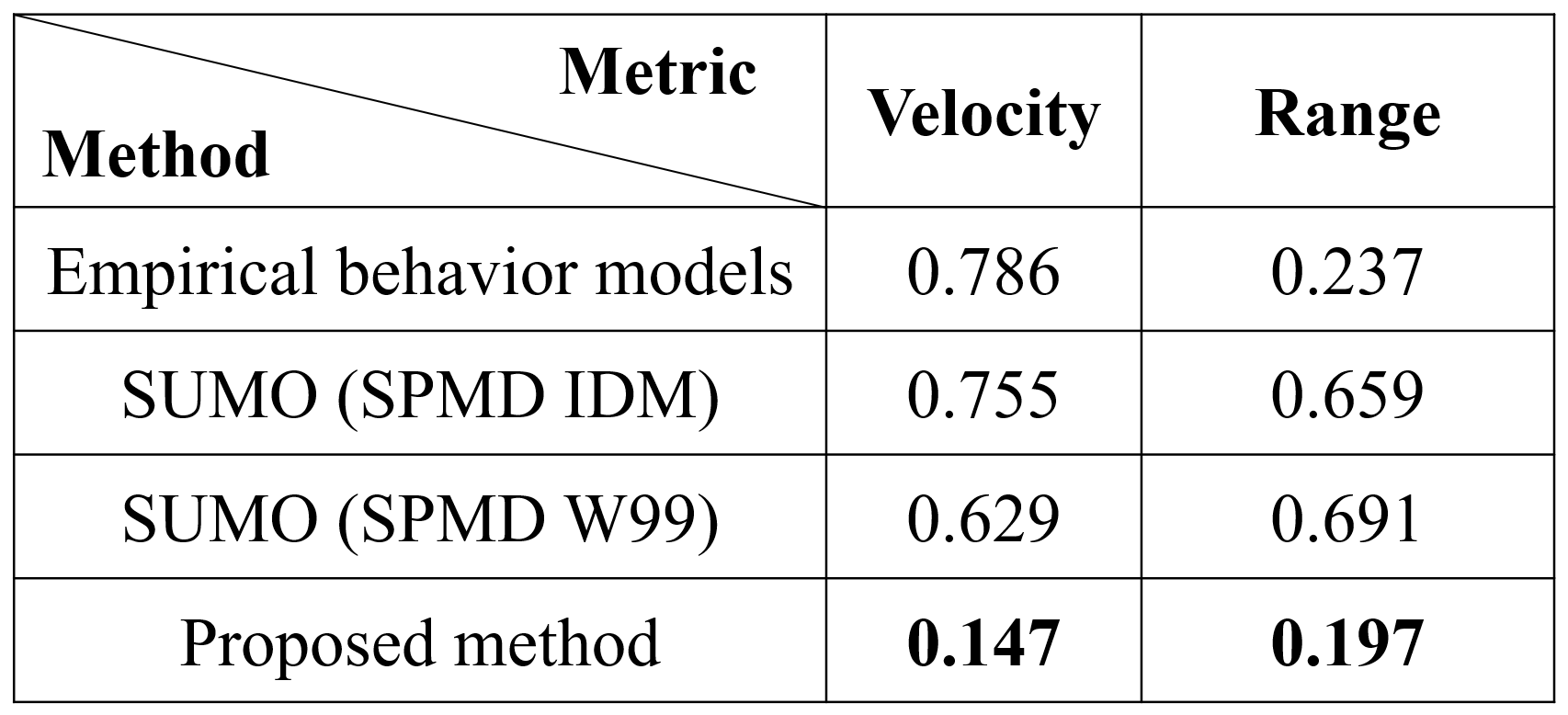}
\par\end{centering}

\end{table}

In addition to the velocity and range distributions, we also calculated
the lane-changing statistic of the proposed NDE to further examine
its lateral behavior performance. From the simulation results, the average travel distance
for one lane change is 4.86 kilometers. In the real-world driving
environment, the same statistic is 4.45 kilometers per lane change
on the highway \cite{lee2004comprehensive-LC-statistics}.
Therefore, the proposed NDE can also reproduce
a reasonable number of lane changes as in the real-world driving environment, which can
further demonstrate the fidelity of the proposed NDE.

\subsection{AV testing using the proposed NDE}


To further demonstrate the capability of the proposed NDE model for AV testing, we 
test the safety performance of an AV model utilizing the proposed NDE and SUMO (SPMD IDM) environments, respectively.  Following previous studies on AV testing \cite{zhao2016accelerated,feng2020part1,feng2020part2,feng2021intelligent-NADE},
the AV accident rate in the NDE is chosen as the measurement, 
and the Monte Carlo method is applied to estimate the accident rate as shwon in Eq. (\ref{eq:AV-performance-expected-probability}). 
Specifically,
the IDM and MOBIL models \cite{feng2021intelligent-NADE} are used as the AV model, 
one simulation test is conducted for a constant driving distance (400 m) of the AV, and the testing result (accident or not) of each simulation is utilized to calculate 
the accident rate. To obtain accurate evaluation results, the AV will run $5\times10^{6}$
simulations in the proposed NDE and SUMO environments, respectively, which is approximately 1.2 million driving miles for the AV under test. The experiments take around four hours using the University of
Michigan's Great Lakes High-Performance Computing cluster with 1,300 cores 
(Intel Xeon Gold 6154 processor) and 9,000 GB RAM. 

\begin{figure}[tbh]
\centering
\subfloat[]{\begin{centering}\label{fig:crash-rate}
\includegraphics[width=0.8\columnwidth]{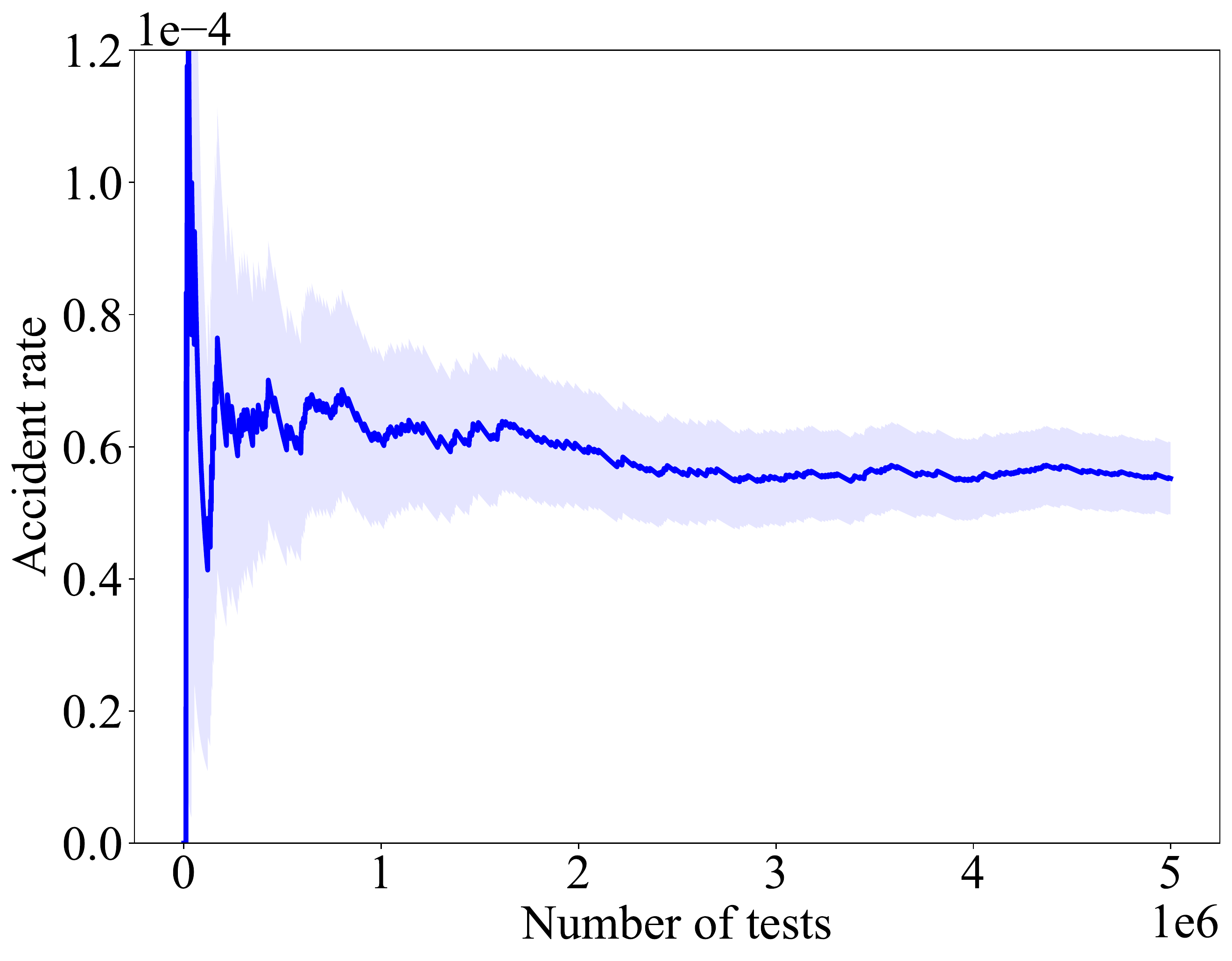}
\par\end{centering}

}



\subfloat[]{\begin{centering}\label{fig:crash-type}
\includegraphics[width=0.8\columnwidth]{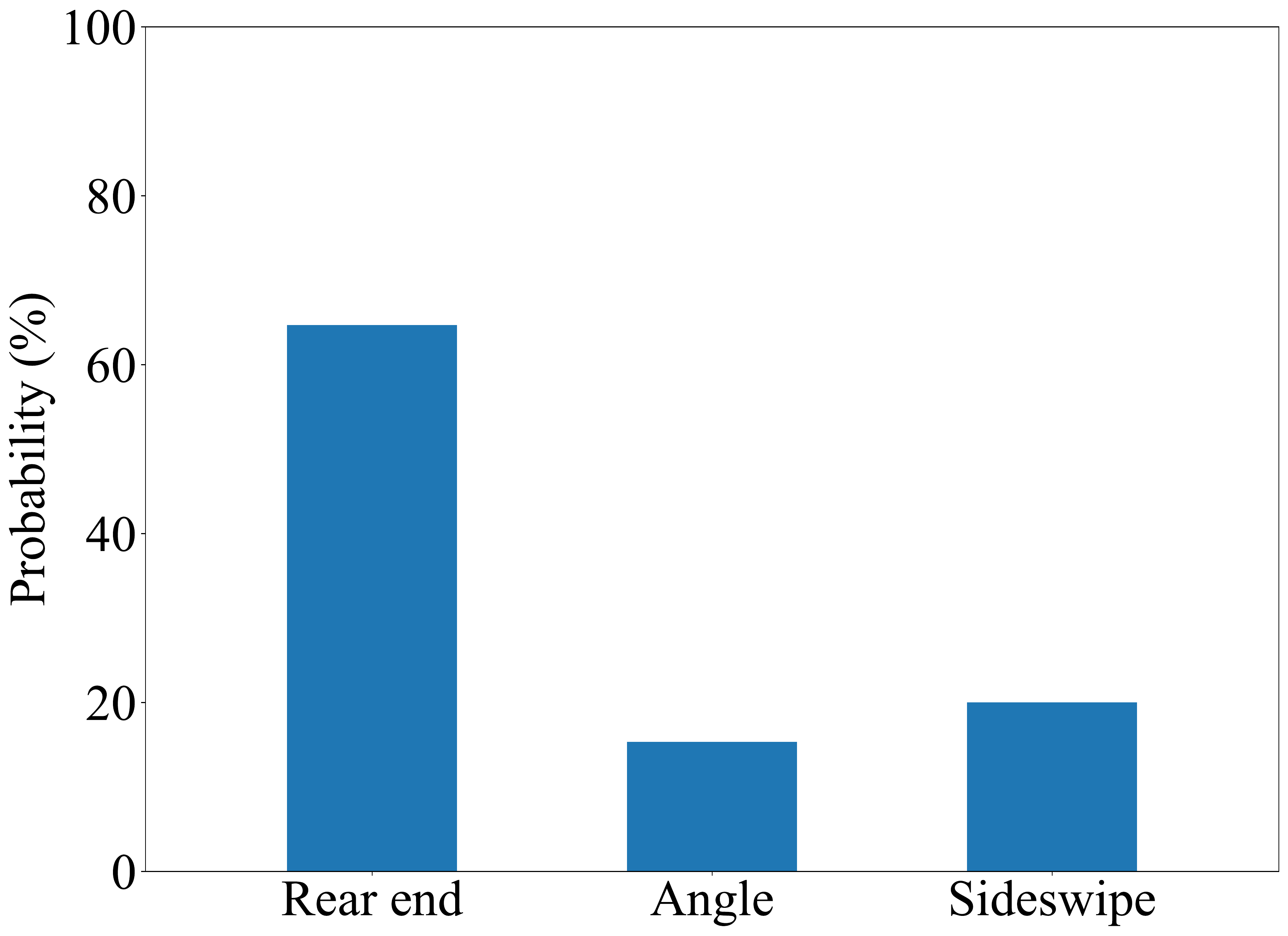}
\par\end{centering}

}

\caption{AV testing and evaluation results using the proposed NDE: (a) estimation results of
the accident rate (shaded area denotes the 90\% confidence interval),
(b) accident type distribution. \label{fig:fig10-AV-testing-results}}

\end{figure}

During the simulation tests in the SUMO environment, there is no accident occurred so the estimated AV accident rate is zero. This is because the SUMO environment is designed for accident-free simulations, and therefore, it cannot effectively evaluate the AV safety performance.
During the simulation tests in the proposed NDE, however, the estimated accident rate of the AV is $5.5\times10^{-5}$ accidents per simulation, where 276 accidents are generated in the AV testing process. The estimation results with the number of tests can be found in Fig. \ref{fig:crash-rate}, where the shaded area denotes the 90\%
confidence interval, and the accident type distribution is shown in Fig. \ref{fig:crash-type} based on the definitions of the National Highway Traffic Safety Administration \cite{FARS-FARS}. The results show that the proposed NDE can successfully generate diverse safety-critical situations to evaluate the AV safety performance. With more distributionally accurate NDE models, the safety performance of AVs could be evaluated more effectively.

\section{Conclusions\label{sec:Conclusions}}

In this paper, we propose a distributionally consistent NDE modeling
framework for AV testing purposes. We first propose to directly generate
empirical behavior models using large-scale NDD. These empirical behavior
models can serve as basic models since they will converge in probability
to the real behavior models when the data size and quality are sufficiently
large and accurate. To address the error accumulation issue and guarantee
the accuracy of the NDE throughout the simulation, a robust modeling
step based on the Markov process is proposed, which optimizes
the empirical models by matching simulated stationary distributions 
with the ground truth. The proposed method is validated for the multilane
highway driving environment using large-scale real-world NDD. The
vehicle speed and range distributions generated by the proposed NDE are consistent with the empirical
ground truth, which are important for AV
testing, and the lane-changing statistic is also examined
to be consistent with the ground truth. Moreover, the generated NDE is utilized to test the safety performance of an AV agent, which further validates the
effectiveness of the proposed NDE.

There are certain limitations in the current work that can be addressed
by future research. For example, homogeneous vehicles are considered in this
study where all background vehicles share the same behavior models
and vehicle classes. In the real-world driving environment, there
are different vehicle classes, such as sedan, truck, heavy vehicle,
etc. Moreover, within a specific vehicle class, they may have different
behavior models, for example, some drivers are more aggressive, and
some drivers are more conservative.
We leave these for future research.

\appendices{}

\section{Naturalistic driving data processing \label{sec:Appendix-NDD-processing}}

First, the original data were segmented into trajectories with the
following requirements: (1) the time should be continuous and does
not have discontinuity of more than 2 seconds, (2) the identification
of the surrounding vehicles is consistent throughout the trajectory,
(3) each trajectory should last more than 3 seconds. Data that are
too noisy (with severe speed and/or position oscillations) are discarded.

\begin{figure}
\begin{centering}
\includegraphics[width=0.75\columnwidth]{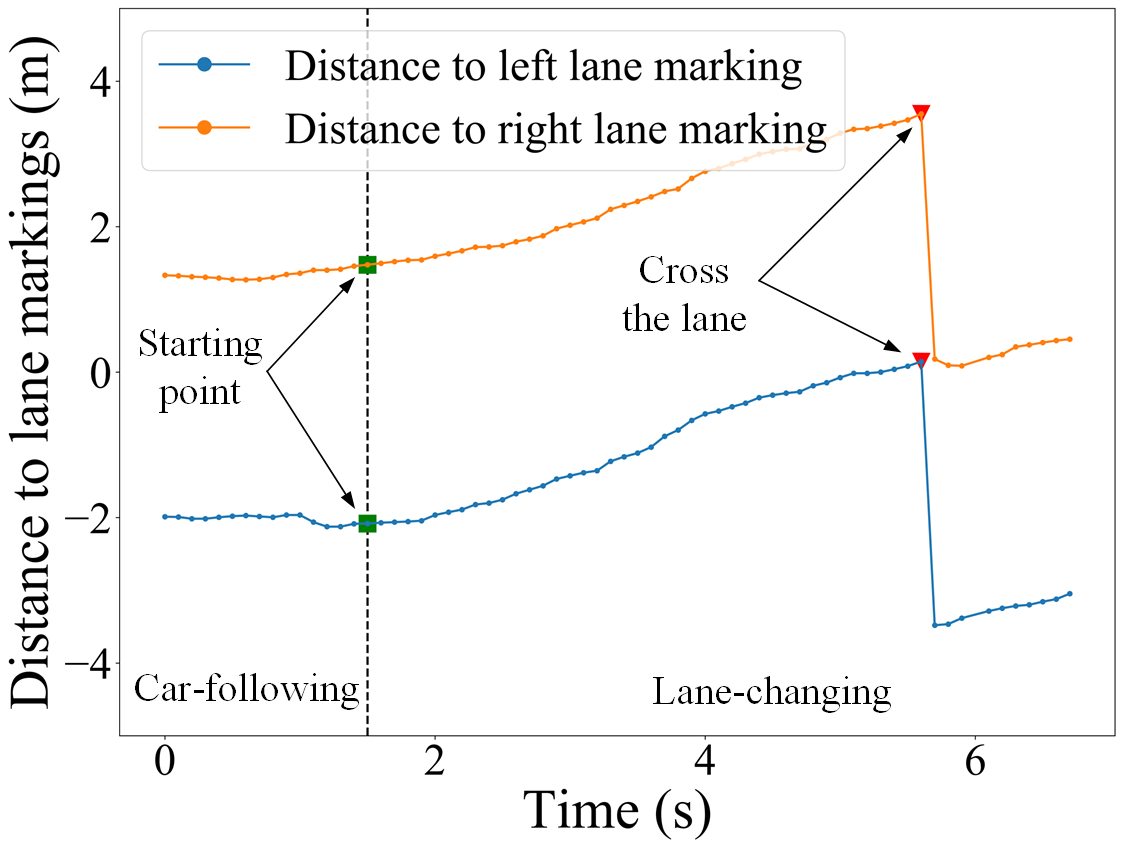}
\par\end{centering}
\caption{Lane-changing events identification example: the subject vehicle is
doing a left lane change. \label{fig:fig5-Lane-changing-events-identificat}}

\end{figure}

Second, considering the driving environment of the subject vehicle, trajectory
data points will be categorized into the six situations as illustrated in Fig. \ref{fig:fig3a-6-categories}, correspondingly.
An important step in the categorization is to identify the lane-changing
event by analyzing the lateral distance to the lane markings. Fig.
\ref{fig:fig5-Lane-changing-events-identificat} shows an example
that the ego-vehicle is doing a left lane change. Note that the sign
of the distance to lane marking differentiates the left and right
lane markings. Fig. \ref{fig:fig5-Lane-changing-events-identificat}
shows the whole process of a left lane-changing event: the vehicle
is approaching the left lane marking and cross the lane when the distance
to the left lane marking equals zero, and then change to the maximum
(lane width) after entering the target lane. The starting moment and
crossing the lane moment of lane-changing events can be identified
based on both the distance and slope change of the distance, accordingly.
Similar techniques are also utilized for lane-changing detection in
a recent study \cite{de2020-Real-world-scenario}. The starting point
of the lane-changing event will be used for calculating the lane change
probability.

Third, for each data point, the state was discretized using the following
resolutions: except for the speed resolution of longitudinal behavior
is 0.2 m/s, the speed and spacing resolutions are 1 m/s and 1 m for
all other situations. The longitudinal acceleration discretization
resolution is set as 0.2 $m/s^{2}$. Therefore, at each decision
moment, a vehicle has 33 possible actions: left lane change, 31 longitudinal
accelerations (range from -4 to 2 $m/s^{2}$ with 0.2 as the discretization
resolution), right lane change. 

Finally, the Moving Average (MA)
\cite{dixon1951-moving-average} smoothing technique was applied.
During the smoothing process, states leading to an inevitable crash
will not be included. After that, the empirical probability of each
action at each state can be calculated by its frequency in the dataset
of the corresponding category. These constitute empirical data-driven
vehicle behavior models.

\section{Experiment settings\label{sec:AppendixB-sim-settings}}

The experiment settings of the proposed method and the SUMO baseline is discussed in the section. For the proposed method, a three-lane highway driving environment is developed based on an open-source highway
traffic simulator \cite{highway-env}. The bicycle model is implemented
to update vehicle states at a 10Hz frequency. All lane-changing maneuvers
are set completed in 1 second. To be consistent with the NDD, the
speed of the NDE simulation is bounded between 20 to 40 m/s, and the
longitudinal acceleration is bounded between -4 to 2 $m/s^{2}$. Since
both the longitudinal and lateral models are data-driven, there are
inevitably states where no NDD is collected. We will use the IDM and
MOBIL model for states not covered by the proposed behavior models.
The IDM model parameters are calibrated from the same NDD. The MOBIL
model parameters are partially from \cite{gong2018evaluation-MOBIL-Calibration}
which are also calibrated using the SPMD dataset. The detailed model
parameters are listed in the next paragraph. The simulation initialization
method determines the initial state of all vehicles, which includes
the position and speed, etc. A realistic initialization is preferred
to shorten the required warm-up time, which can improve the simulation
efficiency. In this study, a data-driven initialization method is
proposed based on NDD and the details can be found in Appendix. \ref{sec:AppendixC-initialization}. Using the proposed
initialization method, around 60 vehicles will be initialized for
each simulation. It is approximately 1360 vehicles/hour/lane, which
belongs to level of service (LOS) C for multilane highways \cite{HCM-1994-LOS}.
We ran 100 simulations to mitigate the randomness effect. To fully
examine the error accumulation issue, each simulation ran 15 minutes,
which includes 10 minutes of warm-up time and 5 minutes of data collection.

The calibrated parameters of the IDM model using the SPMD are: maximum
acceleration (0.8 $m/s^{2}$), desired speed (37 m/s), exponent parameter
(3.0), comfortable deceleration (-1.3 $m/s^{2}$), gap at standstill
(0.1 m), and desired time headway (0.8 s). To generate a stochastic
version IDM, the output acceleration will follow a Gaussian distribution
with the original output acceleration as mean and 0.3 as standard
deviation. The parameters of the MOBIL model are: politeness factor
(0.1), utility threshold (0.2 $m/s^{2}$), and maximum safe deceleration
(-3 $m/s^{2}$).

For the SUMO baseline, all simulation settings are the same as the proposed method. Each simulation lasts for 15 minutes, which includes 10 minutes of warm-up time and 5 minutes of data collection. We set the input traffic flow as 1360 vehicles/hour/lane and 60 vehicles will be generated in each simulation. The initial speed of the vehicle is set as 32 m/s based on NDD. To account for the stochasticity of the simulator, we ran 100 simulations for each model to collect data.

\section{The proposed data-driven simulation initialization method\label{sec:AppendixC-initialization}}

We propose a data-driven initialization method to sequentially determine
the $(i+1)$-th vehicle state $(x_{i},y_{i},v_{i})$, i.e., longitudinal
position, lateral position, and velocity, of downstream vehicles based
on its upstream i-th vehicles\textquoteright{} states. The lateral
position $y_{i}$ is the same as the lateral coordinates of its lane
center. The first vehicle of each lane is determined by sampling its
longitudinal position inside an initial zone from a uniform distribution
and its velocity from the empirical velocity distribution. This can
be expressed as

\begin{align}
x_{i} & \sim U\left(0,d_{0}\right),\\
v_{i} & \sim\pi_{v}^{\star},
\end{align}
where $U\left(\cdot,\cdot\right)$ denotes the uniform distribution,
$d_{0}$ is the predetermined initial zone size, $v_{0}$ is the speed,
and $\pi_{v}^{\star}$ is the empirical velocity distribution obtained
from the NDD. Based on the state of the $i$-th vehicle, we can determine
its downstream $(i+1)$-th vehicle state. An indicator variable follows
a Bernoulli distribution $I_{i}\sim B\left(p_{CF}\right)$ is sampled
to determine whether the $i$-th vehicle is in the car-following $(I_{i}=1)$
situation or free-driving $(I_{i}=0)$ situation. $p_{CF}$ is the
probability a vehicle is in the car-following case, which can be approximated
from NDD. Then the joint distribution of range $r$ and range rate
$rr$ given the current speed $g(r,rr|v_{i})$ is queried from the
NDD if $I_{i}=1$. Then, the position and velocity of the next (downstream)
vehicle can be obtained by

\begin{align}
x_{i+1} & =x_{i}+r,\\
y_{i+1} & =y_{i},\\
v_{i+1} & =v_{i}+rr,
\end{align}
where $r,rr$ is the range and range rate sampled from $g(r,rr|v_{i})$,
and the lateral position of the $(i+1)$-th vehicle is the same as
the $i$-th vehicle since they are in the same lane. If $I_{i}=0$,
the next vehicle is randomly generated in a new initial zone outside
the car-following observation range, and the velocity follows the
empirical velocity distribution. It can be represented as

\begin{align}
x_{i+1} & =x_{i}+d_{obs}+r,\\
r & \sim U\left(0,d_{0}\right),\\
y_{i+1} & =y_{i},\\
v_{i+1} & \sim\pi_{v}^{\star},
\end{align}
where $d_{obs}$ is the predetermined car-following observation range
suggested by the NDD. Vehicles on each lane can be sequentially generated
by repeating this process. Note that unrealistic initialization that leads to trivial initial collision will be rejected and resampled. Based on the data, the parameters used
in the study are $d_{0}=50m$, $p_{CF}=0.68$, and $d_{obs}=115m$.



\section{Performance of baseline methods using parameters from the literature\label{sec:AppendixE-literature-param}}

\begin{figure}[tbh]
\centering
\subfloat[]{\begin{centering}
\includegraphics[width=0.8\columnwidth]{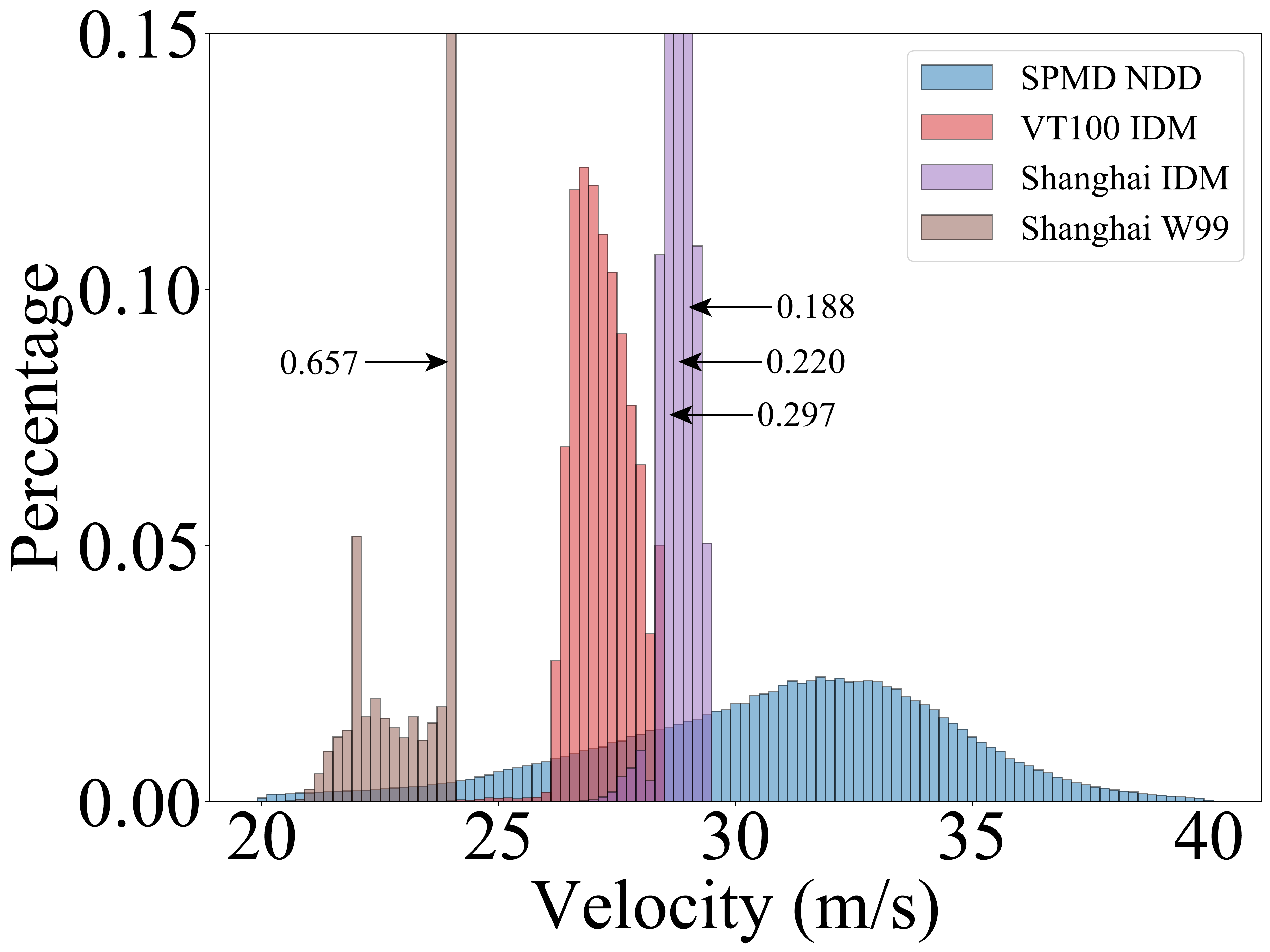}
\par\end{centering}
}

\subfloat[]{\begin{centering}
\includegraphics[width=0.8\columnwidth]{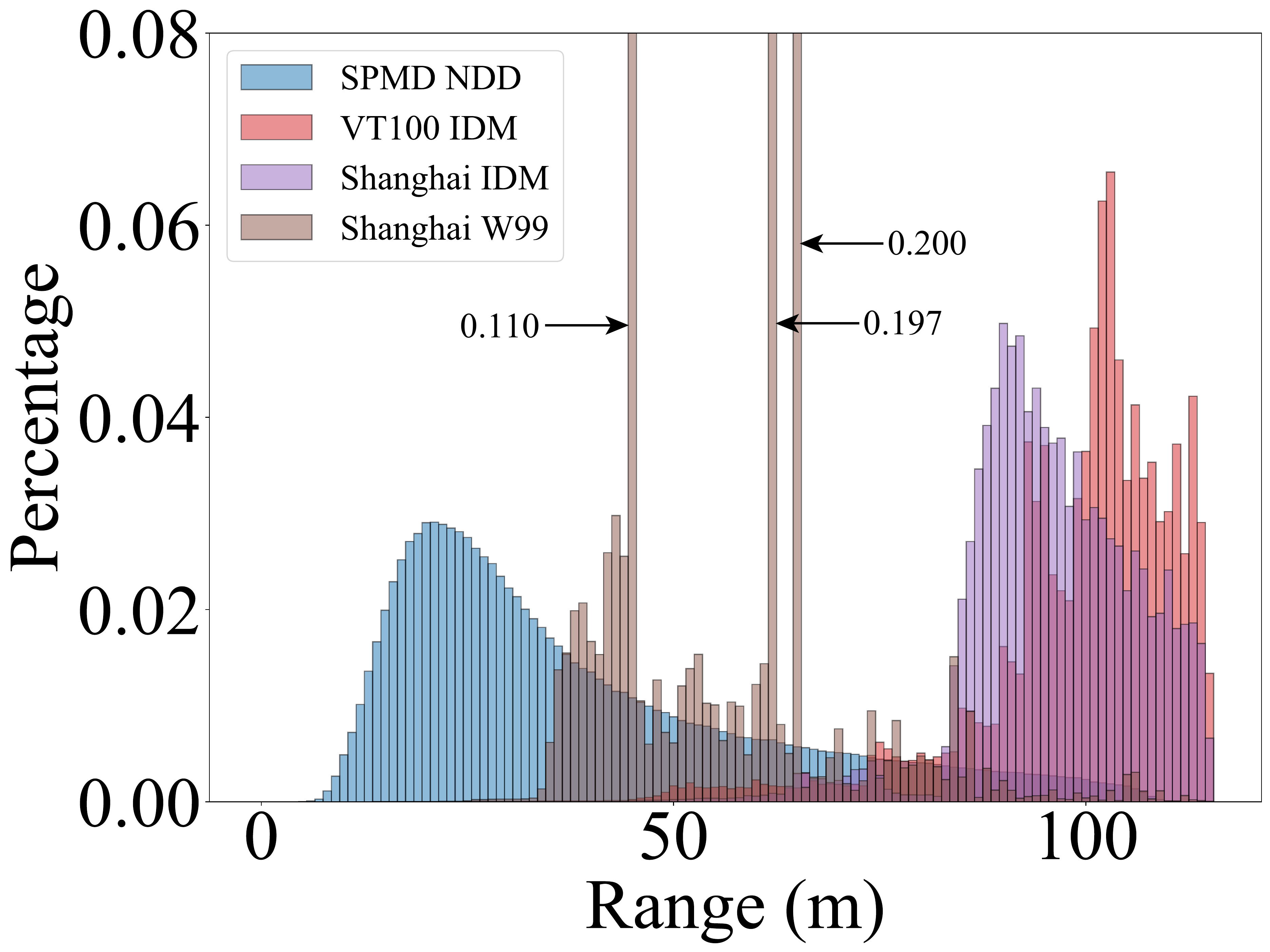}
\par\end{centering}
}

\caption{Velocity and range distributions that generated using model parameters
from the literature. \label{fig:fig11-literature-param}}
\end{figure}

Besides using parameters calibrated by ourselves, we also use model
parameters from the literature to fully illustrate the performance
of existing methods which includes those calibrated by the NDD from
Virginia, US \cite{sangster2013-CF-calibration} (denoted as VT100
IDM), and Shanghai, China \cite{zhu2018-CF-calibration} (denoted
as Shanghai IDM and Shanghai W99). The vehicle initial speed is set
based on the parameters of each model, which is 26 m/s for VT100 IDM,
28 m/s for Shanghai IDM, and 22 m/s for Shanghai W99. Other simulation
settings are the same as discussed in Appendix. \ref{sec:AppendixB-sim-settings}.,
where each model runs 20 simulations to collect data. The results
are shown in Fig. \ref{fig:fig11-literature-param}. Although the
NDD reported in Fig. \ref{fig:fig11-literature-param} is not the
same one used for calibration in the literature, the comparison is
still informative. The distributions generated using existing methods
are concentrated in relatively dense regions, however, the real-world
distribution spreads naturally among the whole intervals.

\section*{Acknowlegment}
The authors would like to thank Dr. Bo Yu
for his help in the data querying process.

\bibliographystyle{IEEEtran}
\bibliography{IEEEabrv,IEEEexample,reference}

\end{document}